\documentclass[namedreferences]{SolarPhysics}
\usepackage[optionalrh,solaenum]{spr-sola-addons} 
\usepackage{graphicx}                    
\usepackage{amssymb}                    
\usepackage{color}                       
\usepackage{url}                         

\begin{document}

\begin{article}

\begin{opening}

\title{A spatio-temporal description of the abrupt changes in the photospheric magnetic and Lorentz-force vectors during the 2011 February 15 X2.2 flare}

%
\author{G.J.D.~\surname{Petrie}      
       }

%

%
  \institute{National Solar Observatory, Tucson, AZ 85719, USA
                     email: \url{gpetrie@noao.edu}\\
             }

\begin{abstract}

The active region NOAA 11158 produced the first X-class flare of Solar Cycle 24, an X2.2 flare at 01:44~UT on 15 February 2011. The \textit{Helioseismic and Magnetic Imager} (HMI) instrument on the \textit{Solar Dynamics Observatory} satellite produces 12-minute, $0.\!\!^{\prime\prime}5$~pixel$^{-1}$ vector magnetograms. Here we analyze a series of these data covering a 12-hour interval centered at the time of this flare. We describe the spatial distributions of the photospheric magnetic changes associated with the flare, including the abrupt changes in the field vector, vertical electric current and Lorentz force vector acting on the solar interior. We also describe these parameters' temporal evolution. The abrupt magnetic changes were concentrated near the neutral line and in two neighboring sunspots. Near the neutral line, the field vectors became stronger and more horizontal during the flare and the shear increased. This was due to an increase in strength of the horizontal field components near the neutral line, most significant in the horizontal component parallel to the neutral line but the perpendicular component also increased in strength. The vertical component did not show a significant, permanent overall change at the neutral line. The increase in field strength at the neutral line was accompanied by a compensating decrease in field strength in the surrounding volume. In the two sunspots near the neutral line the integrated azimuthal field abruptly decreased during the flare but this change was permanent in only one of the spots. There was a large, abrupt, downward vertical Lorentz force change acting on the solar interior during the flare, consistent with results of past analyses and recent theoretical work. The horizontal Lorentz force acted in opposite directions along each side of neutral line, with the two sunspots at each end subject to abrupt torsional forces relaxing their magnetic twist. These shearing forces were consistent with a contraction of field and decrease of shear near the neutral line, whereas the field itself became more sheared as a result of the field collapsing towards the neutral line from the surrounding volume. The Lorentz forces acting on the atmospheric volume above the photosphere were equal and opposite.
\end{abstract}

%

\end{opening}

%
\section{Introduction}
\label{s:introduction} 

Active region (AR) 11158 produced the first X-class flare of Solar Cycle 24, an X2.2 flare at 01:44 UT on 15 February 2011. The \textit{Helioseismic and Magnetic Imager} (HMI) instrument (Schou et al.~2012) on NASA's \textit{Solar Dynamics Observatory} (SDO) satellite (Pesnell et al.~2012) has been observing continuously since March 2010. HMI vector magnetograms covering several days including the time of this flare were released to the community in late 2011 (Hoeksema et al.~2012). In this paper we analyze a series of these data covering a 12-hour interval centered at the time of the X2.2 flare.

SDO/HMI produces full-disk vector magnetograms with $0.\!\!^{\prime\prime}5$ pixels every 12 minutes. The HMI instrument generates filtergrams in six polarization states at six wavelengths on the Fe~{\sc I} 617.3~nm spectral line. From these filtergrams, images for the Stokes parameters, $I$, $Q$, $U$ and $V$ are derived. These are inverted for the magnetic vector components by the Very Fast Inversion of the Stokes Algorithm (VFISV) code (Borrero et al.~2010). The $180^{\circ}$ azimuthal field ambiguity is resolved using the ``minimum energy'' method (Metcalf~1994, Leka et al.~2009). The vector magnetograms specially released in late 2011 were derived by the HMI team from the HMI 720-second Stokes-parameter data series, running the VFISV inversion code with very strict convergence criteria to determine the optimal values, using more computing resources than the HMI pipeline is able to use routinely. The $180^{\circ}$ ambiguity in the azimuthal field was computed with spherical geometry in a limited region with generous thresholds on field strength and very gradual annealing,  also requiring significant additional computing resources. In this paper we use this data set to describe the abrupt and permanent field changes that occurred during the flare and characterize the associated Lorentz force vector changes near the main neutral line of the region and within the neighboring sunspots. (Here a change is deemed ``permanent'' if its effects last at least several hours after the flare.)

After many decades of searching for evidence of flare-related field changes in the photosphere (e.g., Severny~1964, Zirin and Tanaka~1981), abrupt, permanent photospheric field changes have been observationally linked to flares in the past two decades (Sudol and Harvey~2005). Wang et al.~(1992, 1994) found rapid and permanent field changes in flaring active regions, but a number of later studies produced inconclusive results; see the discussion in Wang~(2006).  Kosovichev and Zharkova~(1999) reported a sudden decrease in magnetic energy near an X-class flare, during its impulsive phase.  Then, Kosovichev and Zharkova~(2001) reported on regions of permanent decrease of longitudinal magnetic flux in the vicinity of the magnetic neutral line near the 14 July 2001 ``Bastille Day'' flare and linked the change in flux to the release of magnetic energy.  Wang and Liu~(2010) studied 11 X-class flares for which vector magnetograms were available, and found in each case an increase of transverse field at the polarity inversion line. Wang et al.~(2012) and Sun et al.~(2012) analyzed the HMI data for the 15 February 2011 X2.2 flare, the same data set studied in the present paper, and found similar behavior.  These observations support the coronal implosion interpretation (Hudson 2000, Hudson, Fisher and Welsch 2008, Fisher et al.~2012) where, after a coronal magnetic eruption, the remaining coronal field contracts downward resulting in the field become more horizontal at the photospheric level. Fletcher and Hudson~(2008) have given the only detailed explanation so far of how a coronal event could permanently alter a photospheric field.

Distinctive patterns of behavior have also been found in the behavior of sunspot magnetic fields during flares. Parts of the outer penumbral structures decay rapidly after many flares, while neighboring umbral cores and inner penumbral regions become darker (Wang et al.~2004, 2005, Deng et al.~2005, Liu et al.~2005, Wang et al.~2009).  Transverse fields have been found by these authors to decrease in the regions of penumbral decay and to increase at the flare neutral lines.  Li et al.~(2009) found that during the 13 December 2006 X3.4 flare the mean inclination angle of the magnetic field increased in the part of the penumbra that decayed, whereas the inclination angle decreased in the part of the penumbra that was enhanced during the flare and near the magnetic neutral line. The magnetic twist of sunspot fields has been observed to decrease abruptly as a result of flares (Ravindra et al.~2011, Inoue et al.~2011). Gosain et al.~(2009) found coherent lateral motion of the penumbral filaments near the neutral line using high resolution \textit{Hinode} $G$-band images of the 13 December 2006 X3.4 flare, and speculated that these motions were due to impulsive horizontal Lorentz force changes. Gosain and Venkatakrishnan~(2010) investigated the evolution of the vector field during the 13 December 2006 X3.4 flare and found that, in the penumbra of the main sunspot, the observed field was more inclined than the equivalent potential field, and the difference between the observed and potential fields steadily increased before the flare, abruptly decreased during the flare and steadily increased again after the flare. AR~11158 featured much sunspot evolution during the time of the 15 February 2011 X2.2 flare (Liu et al.~2011, Jiang et al.~2012) but the associated magnetic field changes have not been studied in detail. We will do so in this paper, relating the sunspot field changes to those near the main neutral line and calculating the associated changes in the Lorentz force vector. 

Most of the studies described above have focused on field changes near neutral lines or in sunspots. Sudol and Harvey~(2005) and Petrie and Sudol~(2010) adopted a more general approach. Using one-minute the National Solar Observatory's Global Oscillations Network Group (GONG) longitudinal (line-of-sight) magnetograms, Sudol and Harvey~(2005) characterized the spatial distribution, strength and rate of change of permanent field changes associated with 15 X-class flares.   By carefully co-registering the images they succeeded in tracing the field changes pixel by pixel and were able to  show the spatial structure of the changes. They found that the majority of field changes occurred in regions where the field strength reached hundreds of Gauss which, given the $2.\!\!^{\prime\prime}5$~pixel$^{-1}$ resolution of the data, suggests locations close to or within sunspots.  Building on Sudol and Harvey's work, Petrie and Sudol~(2010) analyzed one-minute GONG longitudinal magnetograms covering 77 flares of GOES class at least M5 and found some statistically significant correlations in the field changes. Exploring the relationship between increasing/decreasing longitudinal fields and azimuth and tilt angles at various positions on the disk, they noted that increasing/decreasing longitudinal fields do not correspond straightforwardly to decreasing/increasing changes in tilt angles. However, the overall distributions of longitudinal increases and decreases at different parts of the disk was found to be consistent with Hudson, Fisher and Welsch's~(2008) loop-collapse scenario.

The 15 February 2011 X2.2 flare has already been studied in several papers using a variety of observations and methods. As mentioned above, Wang et al.~(2012) analyzed the HMI data for the 15 February 2011 X2.2 flare, the same data set studied in the present paper, and found an increase of transverse field and field inclination at the polarity inversion line (see also Gosain~2012, Sun et al.~2012). Sun et al.~(2012) calculated nonlinear force-free field models for the coronal field from the HMI vector measurements and argued that the increase in magnetic shear observed at the photosphere is localized at low heights and the shear decreases above a certain height in the corona. Jiang et al.~(2012) described the complex sunspot motions seen around the time of the flare in $G$-band images from the \textit{Solar Optical Telescope} on the \textit{Hinode} satellite and continuum intensity images from SDO/HMI, with particular emphasis on the clockwise motion of the positive sunspot neighboring the main neutral line. Beauregard, Verma and Denker~(2012) measured the horizontal proper motions with local correlation tracking using HMI continuum images and longitudinal magnetograms, and found shear flows along the main neutral line of a few 100~m~s$^{-1}$. Schrijver et al.~(2011) used multi-wavelength data from the \textit{Atmospheric Imaging Assembly} (AIA) on SDO with high spatial and temporal resolution to analyze expanding loops from a flux-rope-like structure over the shearing polarity-inversion line between the central $\delta$-spot groups of AR 11158 that eventually formed a coronal mass ejection moving into the inner heliosphere. Gosain~(2012) used AIA observations to study the evolution of the coronal loops in the region and identified three distinct phases of the coronal loop dynamics during this event: a slow rise phase, a collapse phase and an oscillation phase. In this paper we will focus on characterizing in detail the vector field changes that occurred during the flare by analyzing the three field components and the associated electric currents and Lorentz force changes over a ten-hour interval centered at the start time of the flare.

The paper is organized as follows. In Section~\ref{s:magch} we will present the vector fields observed by HMI before and after the main flare-related field changes took place. We will describe the differences between these vector fields in each spatial dimension and plot the vector field's evolution in time. We will describe in Section~\ref{s:current} the associated electric current changes that occurred during the flare. We will derive the accompanying Lorentz force changes in Section~\ref{s:lorentzfch} and discuss the likely causes of the changes. We will conclude in Section~\ref{s:conclusion}.

\section{The magnetic field vector changes}
\label{s:magch}

\begin{figure} 
\begin{center}
\resizebox{0.99\textwidth}{!}{\includegraphics*[80,150][560,560]{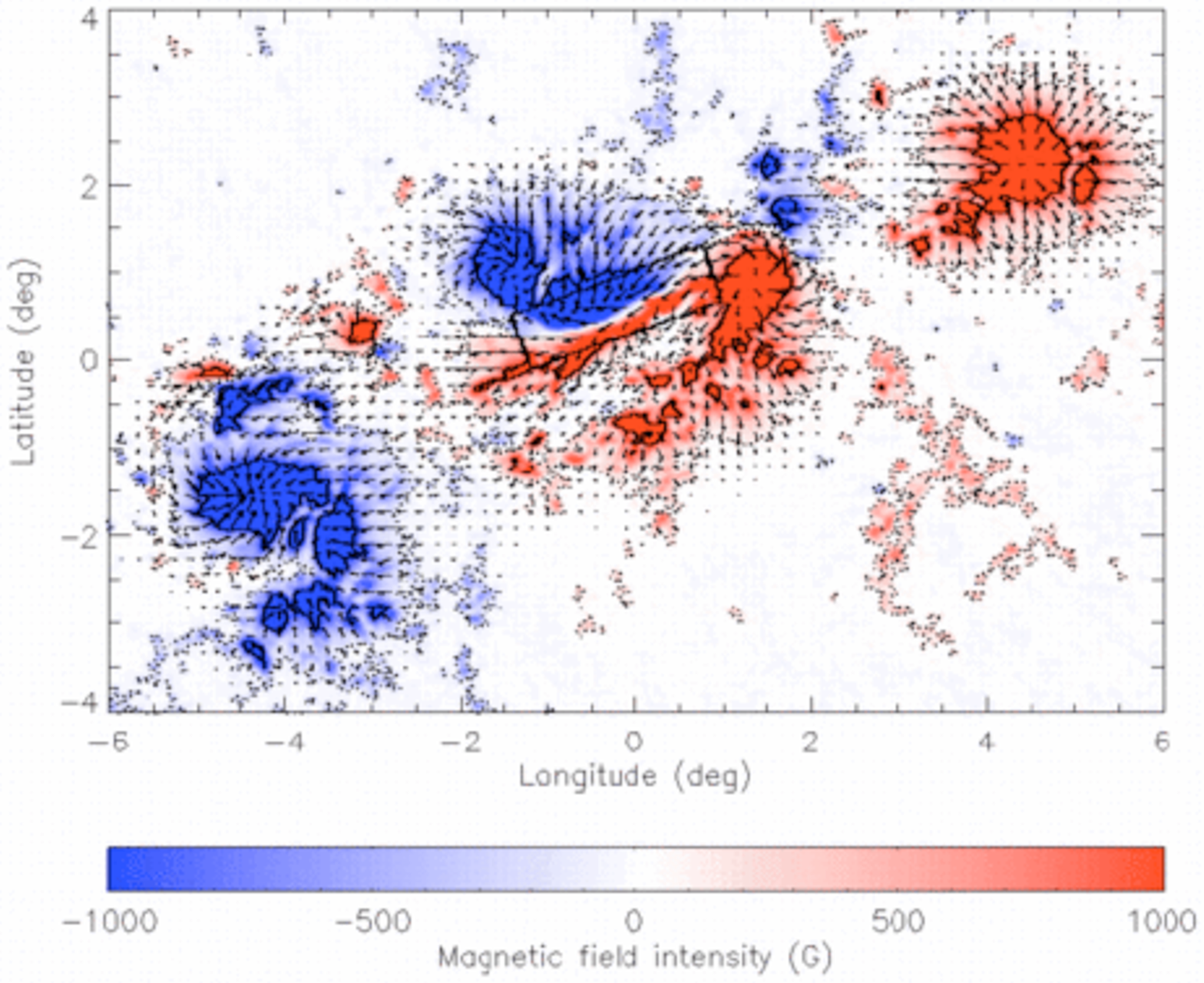}}
\resizebox{0.99\textwidth}{!}{\includegraphics*[80,170][560,560]{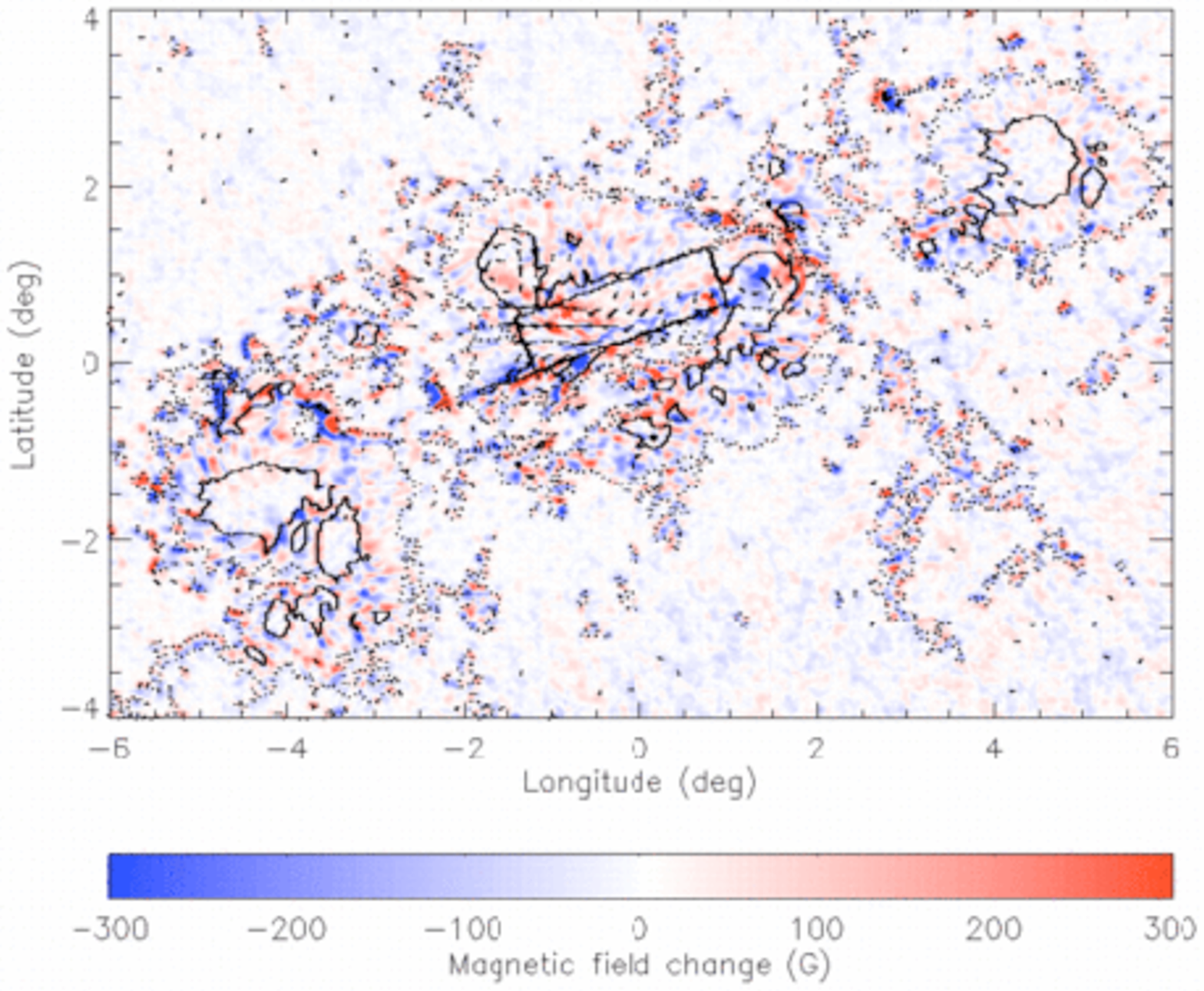}}
\end{center}
\caption{Shown are the vector magnetic field before the flare (top) and the vector field changes (bottom). The vertical components, $B_r$ and $\delta B_r$, are indicated by the color scale and the horizontal components by the arrows, with saturation values $\pm 1000$~G in the top plot and $\pm 300$~G in the bottom plot. Red/blue coloring represents positive/negative vertical field and field changes. The black rectangles mark the region near the neutral line that is used in subsequent analysis. The solid and dotted contours indicate strong ($\mathrm{|}B_r\mathrm{|}>1000$~G) and quite strong  ($\mathrm{|}B_r\mathrm{|}>100$~G) fields, respectively. }
\label{fig:brdbr}
\end{figure}

The X2.2 flare began in NOAA AR~11158 on 15 February 2011 at 01:44~UT when the region was visible on the solar disk at about $20^{\circ}$ south and $10^{\circ}$ west of disk-center. The vector field measurements were released by the HMI team in the form of 12-minute vector magnetogram images $(B_r, B_{\theta} ,B_{\phi})$ in heliographic coordinates $(r, \theta ,\phi )$ on a $600\times 600$ grid with pixel size $0.03^{\circ}$. The top panel of Figure~\ref{fig:brdbr} shows a spatial map of the vertical magnetic field component, $B_r$, before the flare (the image labeled 1:36~UT), with the corresponding horizontal field, $(B_{\phi}, -B_{\theta})$, indicated by arrows. The magnetic field of the region had a complex structure with four major concentrations of intense field. The entire distribution was tilted with respect to the equator. The leading polarity concentration, which was positive, was the most equatorward and the lagging field concentration, which was negative, was the most poleward. So far this arrangement is in line with the Hale-Nicholson law. However, the most interesting part of the active region lay between the leading and lagging field concentrations. There was a bipolar structure composed of a positive leading field concentration and a negative lagging concentration at the same latitude, separated by a highly sheared, S-shaped neutral line that was tilted with respect to the equator at approximately the same angle as the region as a whole. The shear of the field at the neutral line is clearly visible in Figure~\ref{fig:brdbr}, as are the clockwise and anti-clockwise circulations of the field in the negative and positive central field concentrations, respectively. It is in this central portion of the active region that most of the action occurred during the flare, as the bottom panel of Figure~\ref{fig:brdbr} shows. If we have observations of the photospheric vector field at two times, $t=0$ before the field changes begin, and $t=\delta t$ after the main field changes have occurred, the magnetic vector changes due to the flare can be represented by the difference $\delta{\bf B} = {\bf B}(\delta t) - {\bf B}(0)$. The bottom panel of Figure~\ref{fig:brdbr} shows a spatial map of the vertical magnetic field change, $\delta B_r$, with the horizontal field changes $(\delta B_{\phi}, -\delta B_{\theta})$ indicated by arrows. This map was derived by subtracting the 1:36~UT from the 2:00~UT image for each of the three field components. The maps show that the vertical changes were mostly positive/negative on the negative/positive side of the neutral line, weakening the vertical field on both sides of the neutral line, while the horizontal changes point in approximately the same direction as the field itself near the neutral line, strengthening the horizontal field there. Meanwhile the vertical changes in the two central sunspots were mostly positive/negative in the negative/positive spot, weakening the vertical field in both spots. The horizontal changes were anti-clockwise in the negative central spot and clockwise in the positive central spot, weakening the azimuthal field component in each spot.

\begin{figure} 
\begin{center}
\resizebox{0.99\textwidth}{!}{\includegraphics*[80,150][560,560]{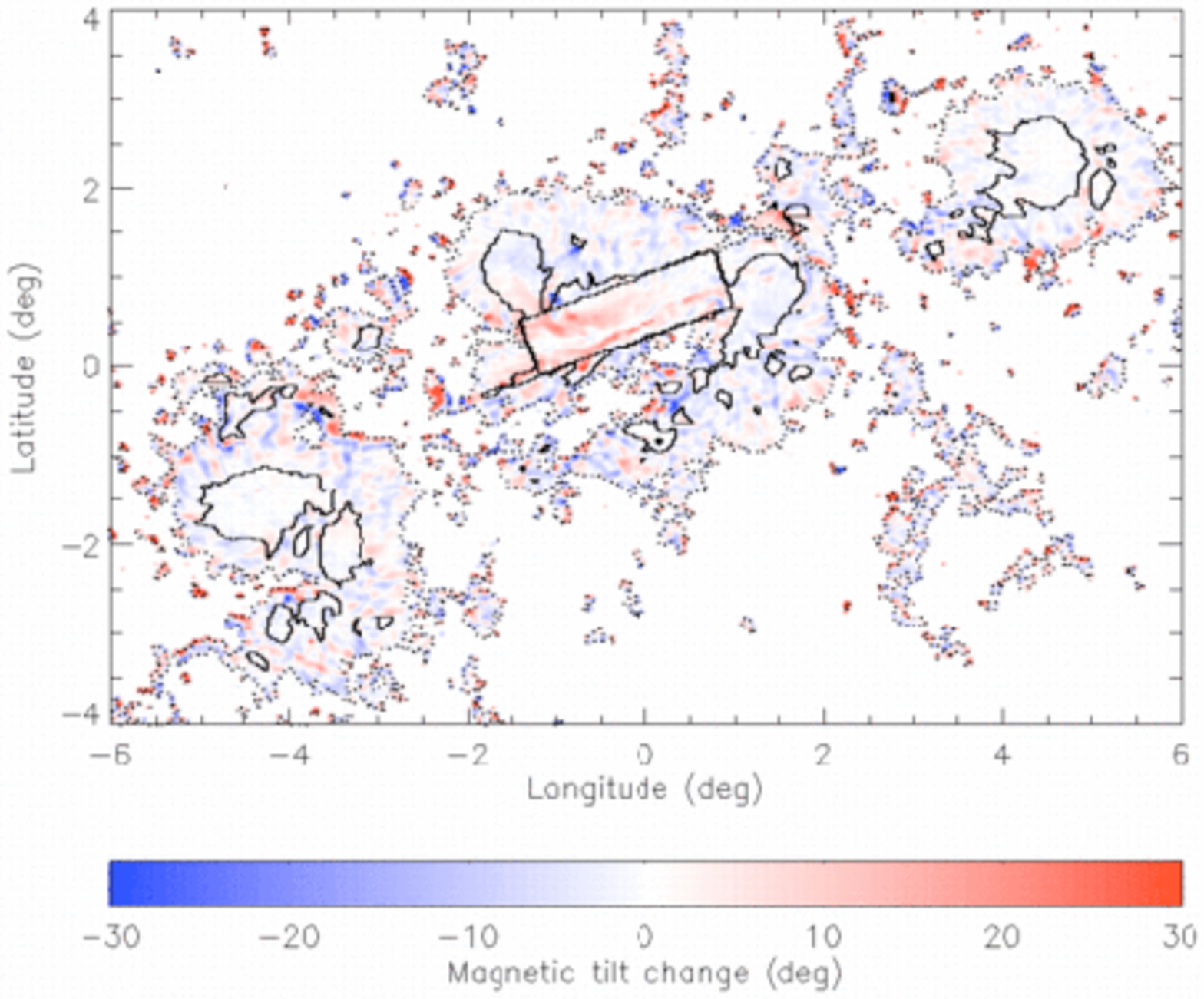}}
\resizebox{0.99\textwidth}{!}{\includegraphics*[80,170][560,560]{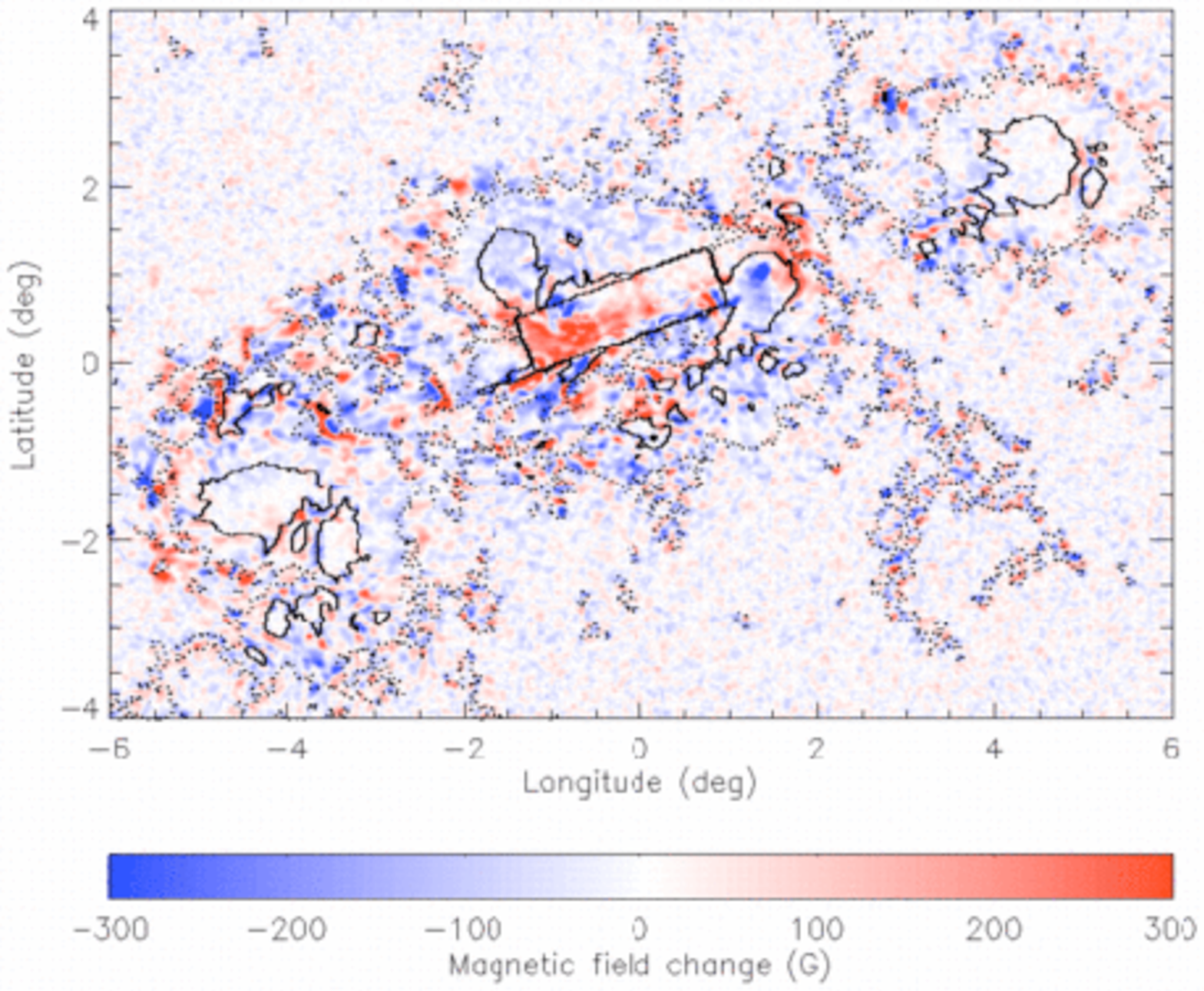}}
\end{center}
\caption{Plotted here are spatial maps of the changes in the field tilt (top) and the total field strength (bottom) during the flare. Red/blue coloring represents positive/negative vertical field change with saturation values $\pm 30^{\circ}$ (top) and $\pm 300$~G (bottom). The black rectangles mark the region near the neutral line that is used in subsequent analysis. The solid and dotted contours indicate strong ($\mathrm{|}B_r\mathrm{|}>1000$~G) and quite strong  ($\mathrm{|}B_r\mathrm{|}>100$~G) fields, respectively. }
\label{fig:dbtilt}
\end{figure}

Figure~\ref{fig:dbtilt} shows spatial maps of the changes in the field tilt $\tan^{-1} ([B_{\theta}^2+B_{\phi}^2]^{1/2}/B_r)$ and the total field strength $B=(B_r^2+B_{\theta}^2+B_{\phi}^2)^{1/2}$ during the flare. There was a clear increase in the tilt angle of the field near the neutral line during the flare, accompanied by a clear increase in total field strength near the neutral line, particularly at the east side of the rectangle in the plot. Also evident is a general decrease in total field strength in the surrounding field outside the rectangle, including much of the field in the two neighboring sunspots. This pattern of increasing tilt and field strength near the neutral line and decreasing tilt and field strength in the surrounding volume is consistent with field abruptly collapsing downwards towards the neutral line and field rushing in from the surrounding volume to fill the resulting void above the neutral line.

\begin{figure} 
\begin{center}
\resizebox{0.99\textwidth}{!}{\includegraphics*[80,150][560,560]{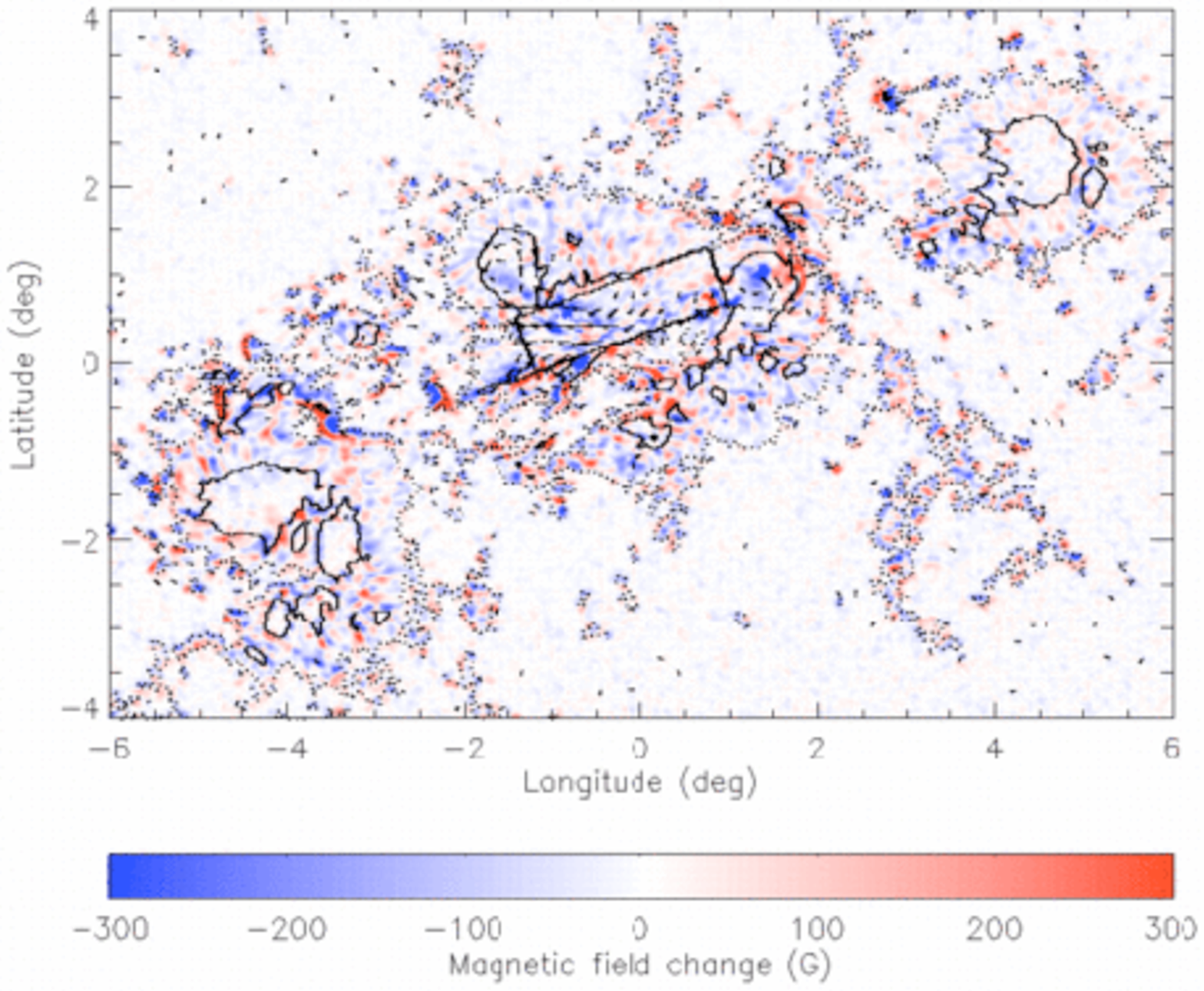}}
\resizebox{0.99\textwidth}{!}{\includegraphics*[80,170][560,560]{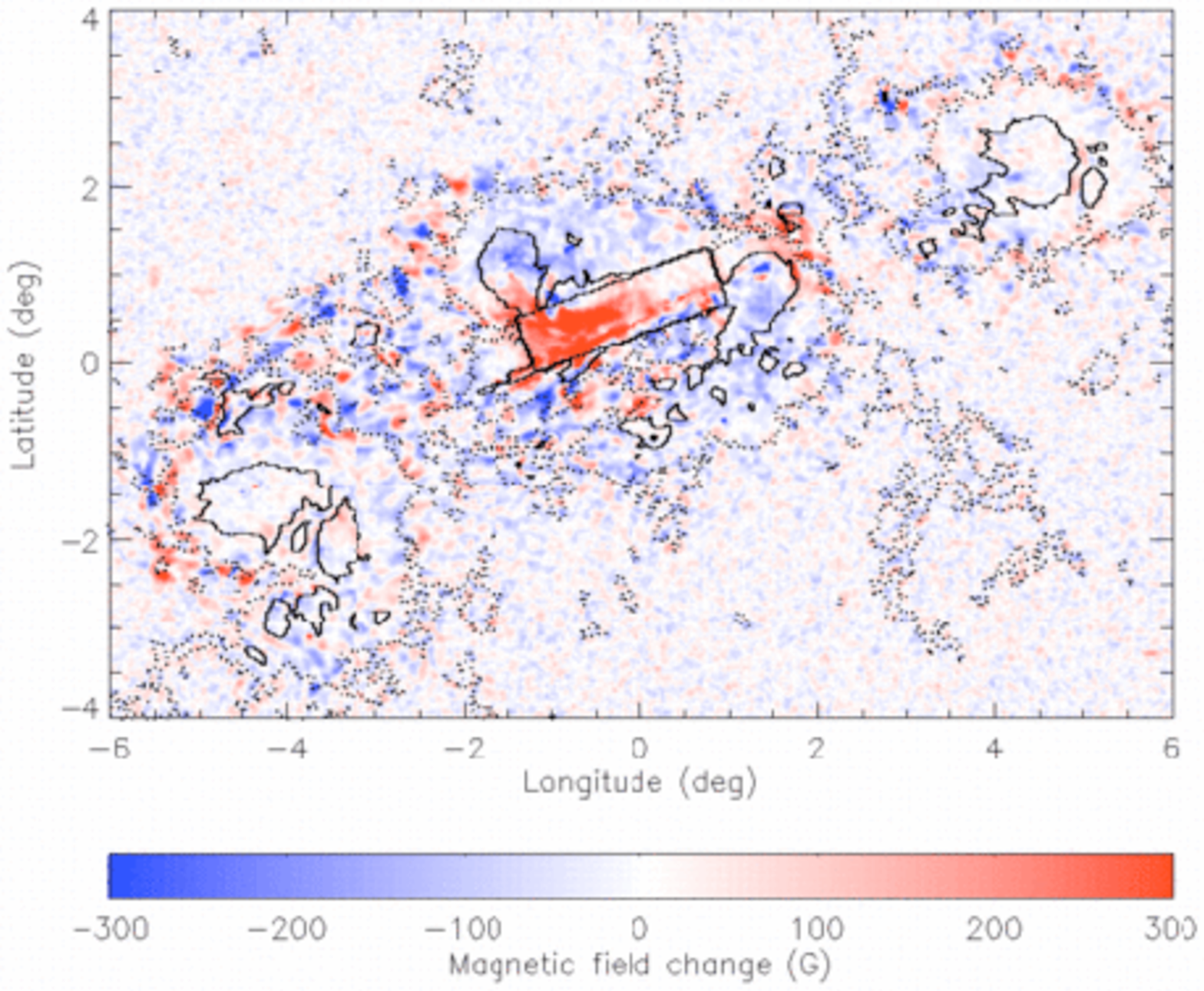}}
\end{center}
\caption{Shown here are spatial maps of the changes in the unsigned vertical field, $\delta \mathrm{|}B_r\mathrm{|}$ (top), and the total horizontal field changes, $\delta B_\mathrm{h}$ (bottom). Red/blue coloring represents positive/negative field change with saturation values $\pm 300$~G. The arrows in the top plot are the same as those in the bottom plot of Figure~\ref{fig:brdbr}. The black rectangles mark the region near the neutral line that is used in subsequent analysis. The solid and dotted contours indicate strong ($\mathrm{|}B_r\mathrm{|}>1000$~G) and quite strong  ($\mathrm{|}B_r\mathrm{|}>100$~G) fields, respectively.}
\label{fig:dbrh}
\end{figure}

Figure~\ref{fig:dbrh} shows spatial maps of the changes in the unsigned vertical field, $\mathrm{|}B_r\mathrm{|}$, and in the horizontal field, $B_\mathrm{h}=(B_{\theta}^2+B_{\phi}^2)^{1/2}$. There is some evidence of weakening of the vertical field during the flare but the distribution of changes is quite complex. The distribution of horizontal field changes is much more striking, with increasing horizontal fields clearly dominating the region near the neutral line surrounded by an area of decreasing horizontal fields. The increase in total field near the neutral line seen in Figure~\ref{fig:dbtilt} was due to the large increase in horizontal field during the flare shown in the bottom panel of Figure~\ref{fig:dbrh}. In fact this horizontal field increase mostly added field in the direction parallel to the neutral line as Figure~\ref{fig:brdbr} shows, and as is clearer in the corresponding temporal plots discussed below.

\begin{figure} 
\begin{center}
\resizebox{0.8\textwidth}{!}{\includegraphics*[20,250][600,550]{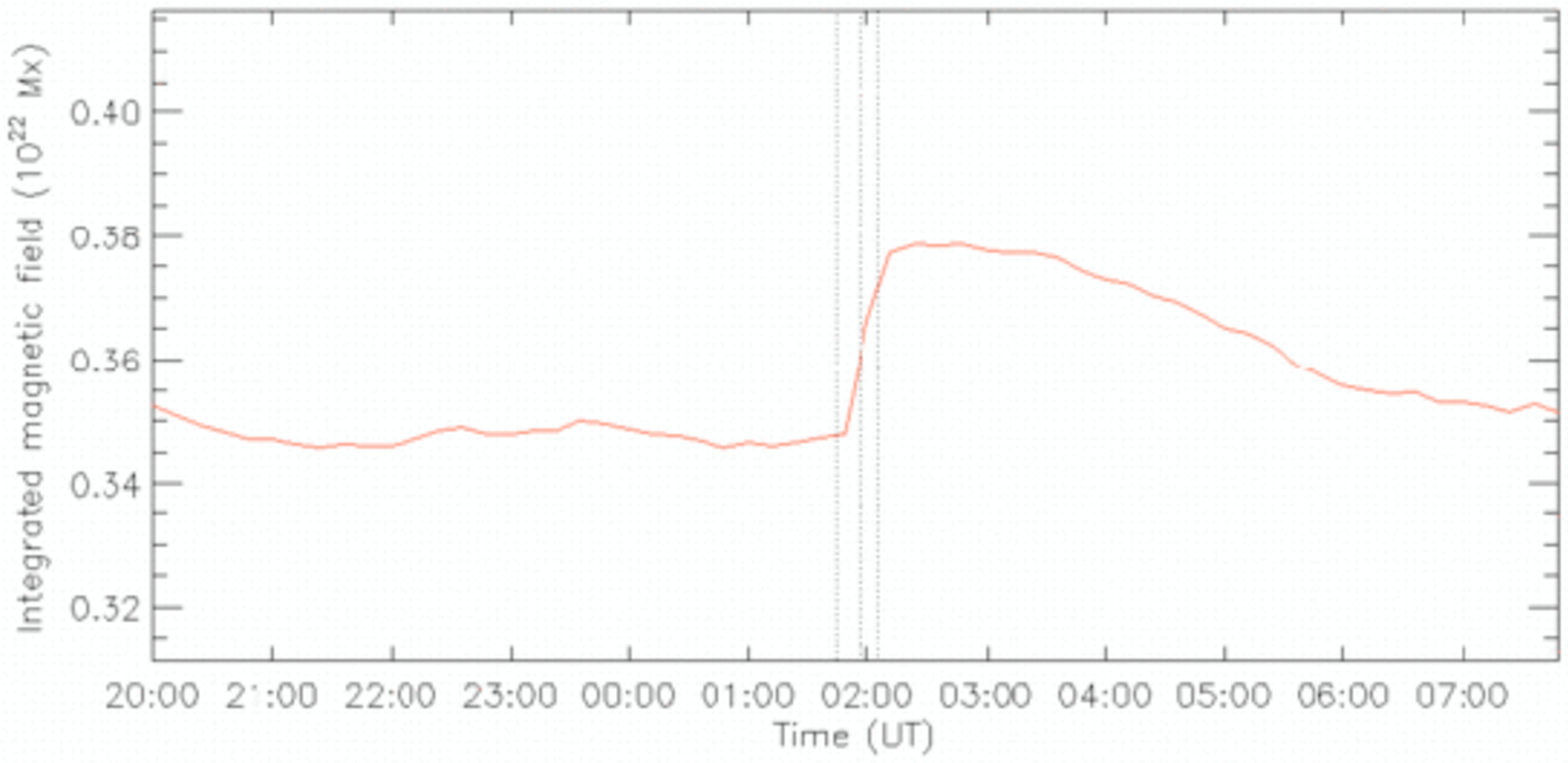}}
\resizebox{0.8\textwidth}{!}{\includegraphics*[20,250][600,550]{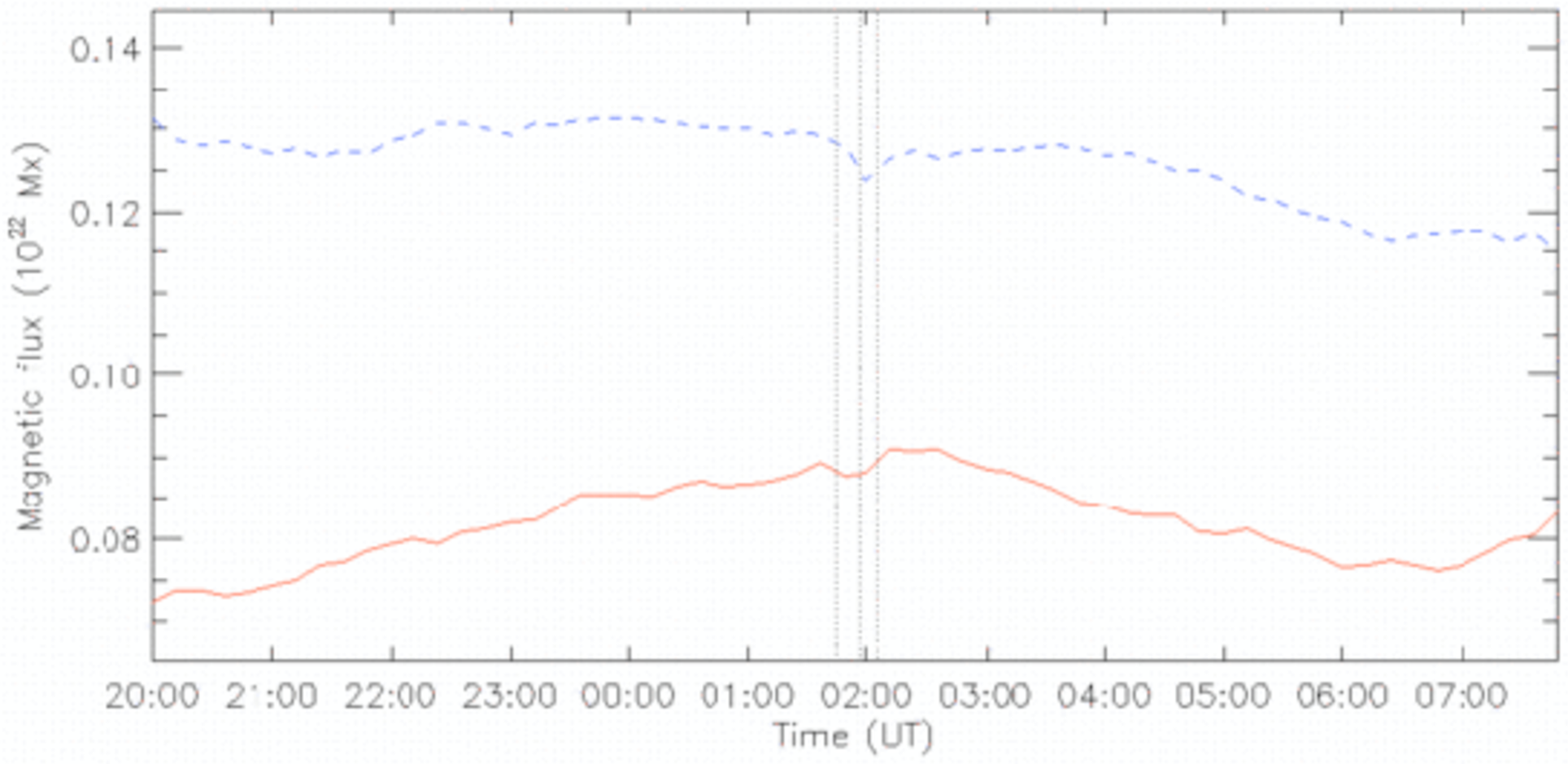}}
\resizebox{0.8\textwidth}{!}{\includegraphics*[20,250][600,550]{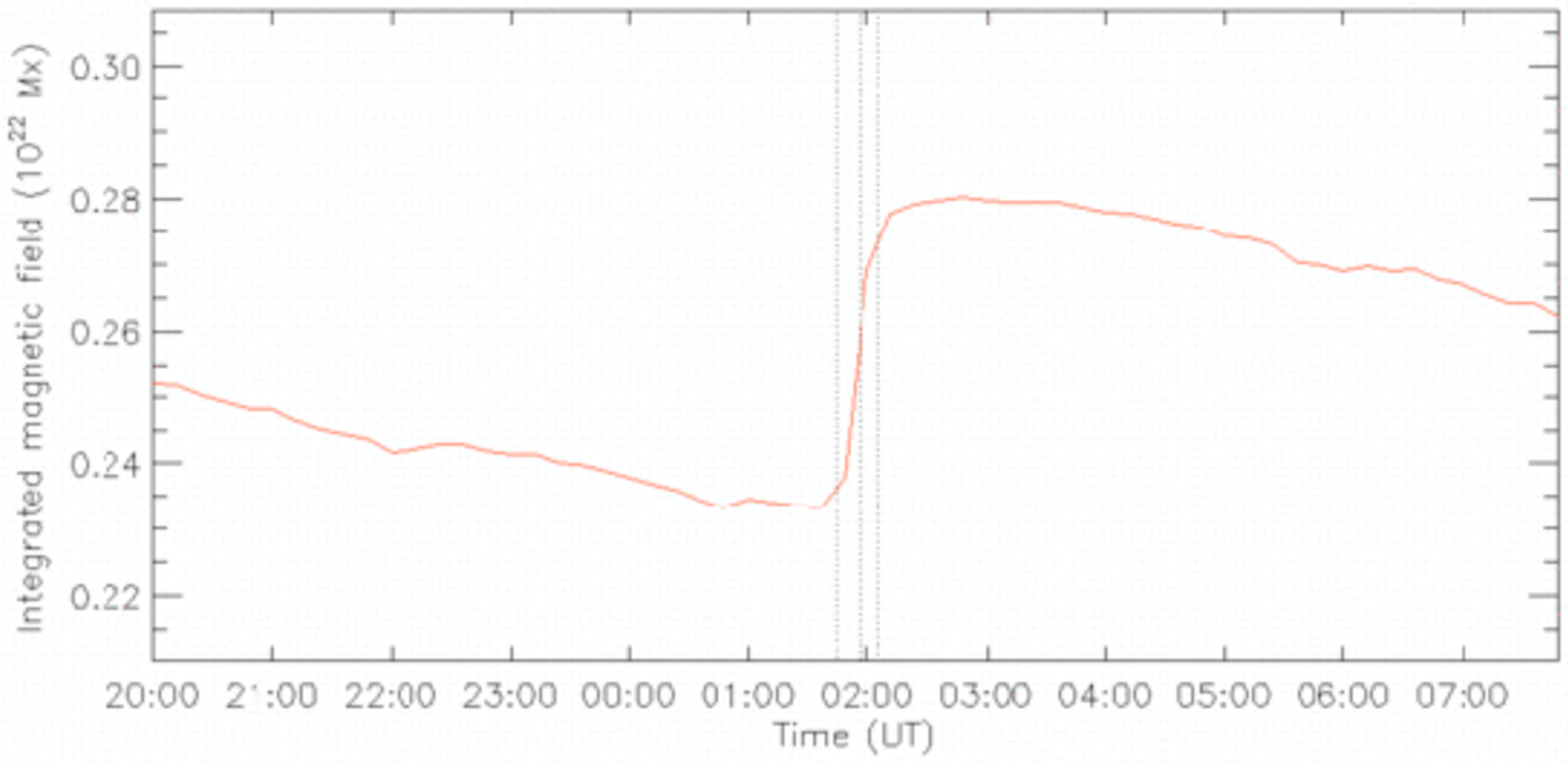}}
\end{center}
\caption{Shown here are the integrated magnetic field strength $B^\mathrm{NL}$ (top), the total vertical flux $B_r^\mathrm{NL}$, (middle) and the integrated horizontal field $B_\mathrm{h}^\mathrm{NL}$, (bottom) near the neutral line are plotted as functions of time. The red/blue solid/dashed lines represent positive/negative field. The area of integration is indicated by the rectangle in Figure~\ref{fig:brdbr}. The vertical lines represent the GOES flare start, peak and end times. }
\label{fig:fnlt1}
\end{figure}

\begin{figure} 
\begin{center}
\resizebox{0.99\textwidth}{!}{\includegraphics*[20,250][600,550]{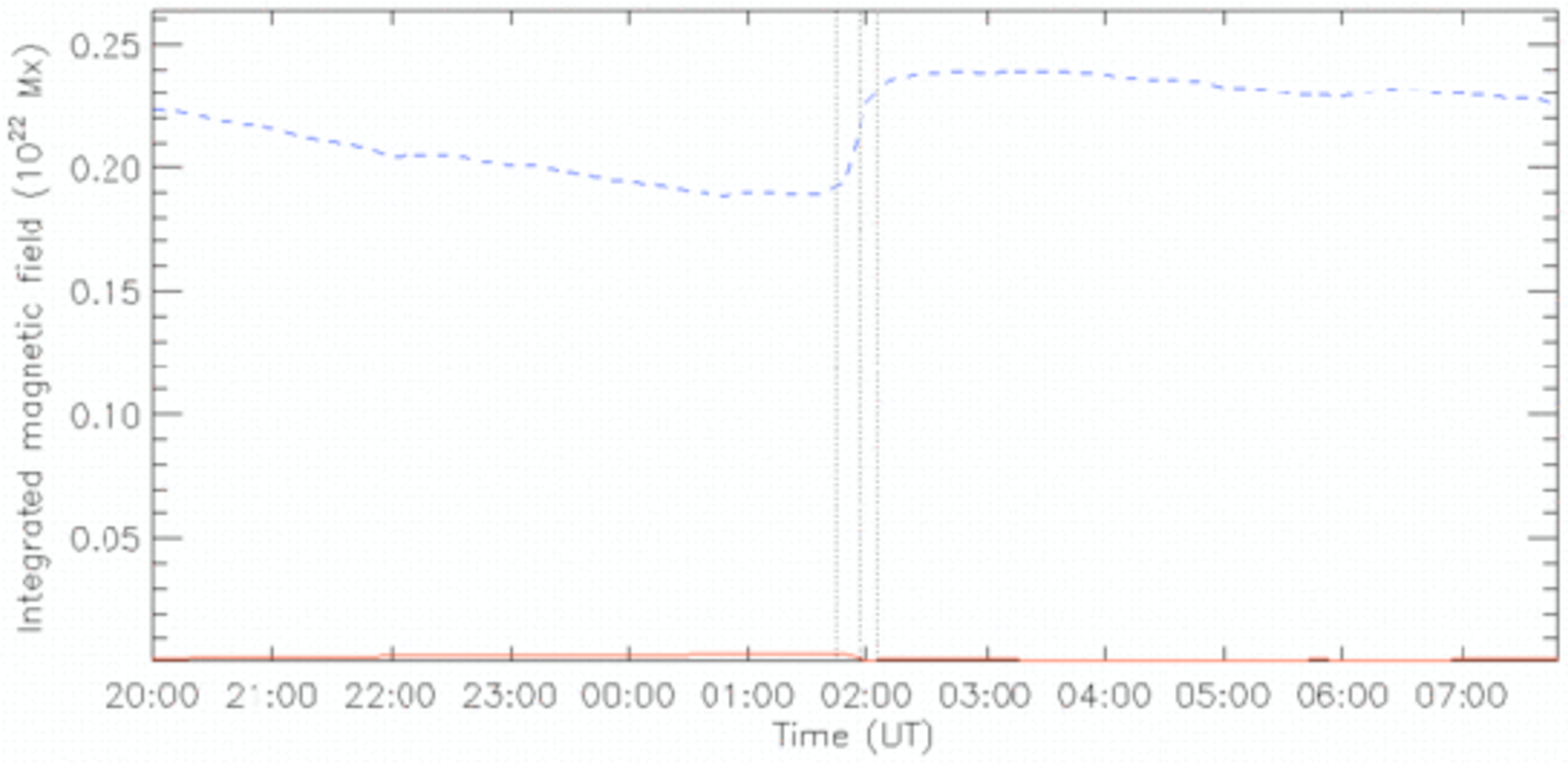}}
\resizebox{0.99\textwidth}{!}{\includegraphics*[20,250][600,550]{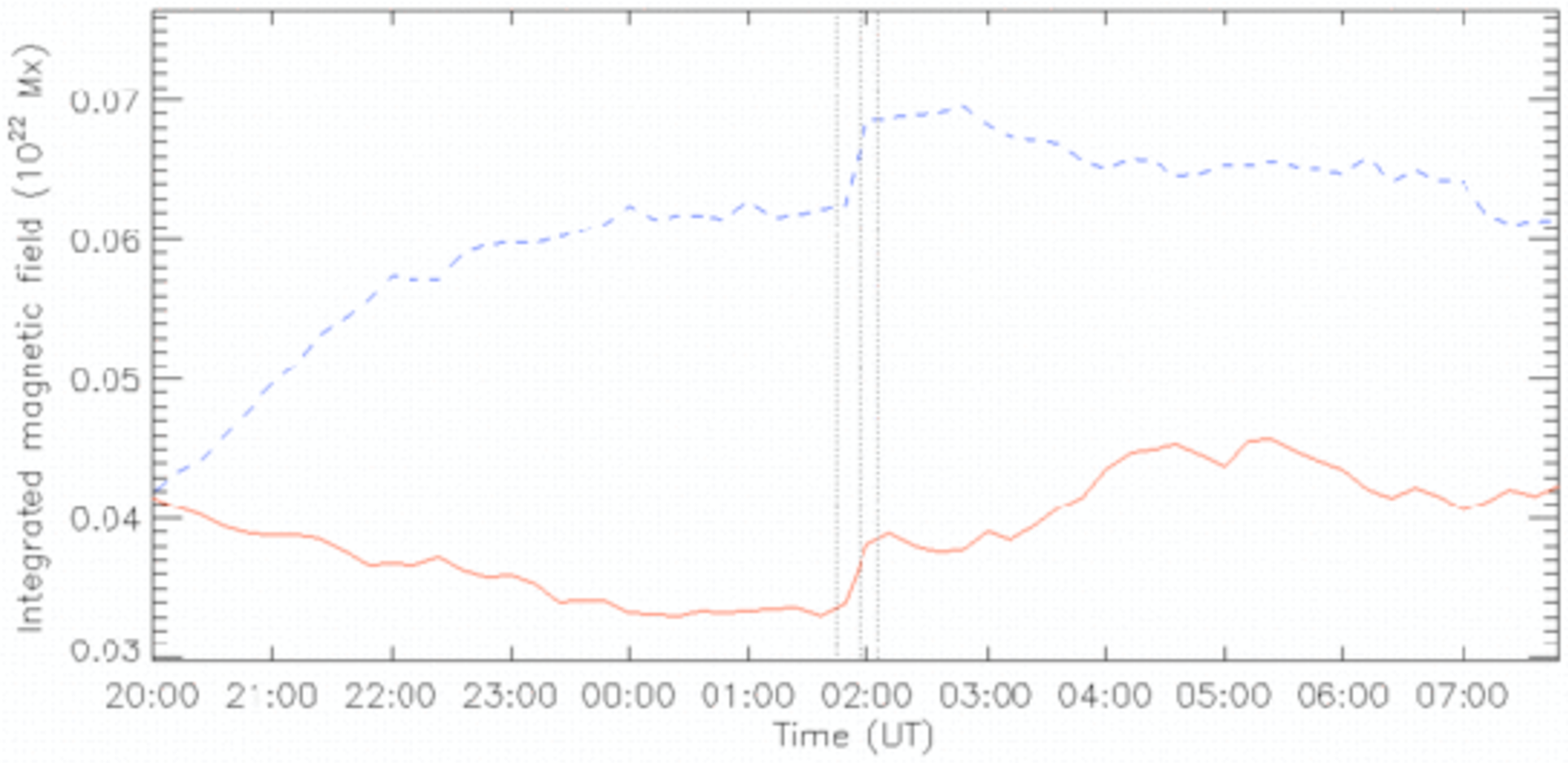}}
\end{center}
\caption{The integrals $B_{\parallel}^\mathrm{NL}$ (top) and $B_{\perp}^\mathrm{NL}$ (bottom) of the horizontal field components parallel and perpendicular to the neutral line are plotted as functions of time. The red/blue solid/dashed lines represent positive/negative field. The area of integration is indicated by the rectangle in Figure~\ref{fig:brdbr}. The vertical lines represent the GOES flare start, peak and end times.}
\label{fig:fnlt2}
\end{figure}

\begin{figure} 
\begin{center}
\resizebox{0.8\textwidth}{!}{\includegraphics*[20,250][600,550]{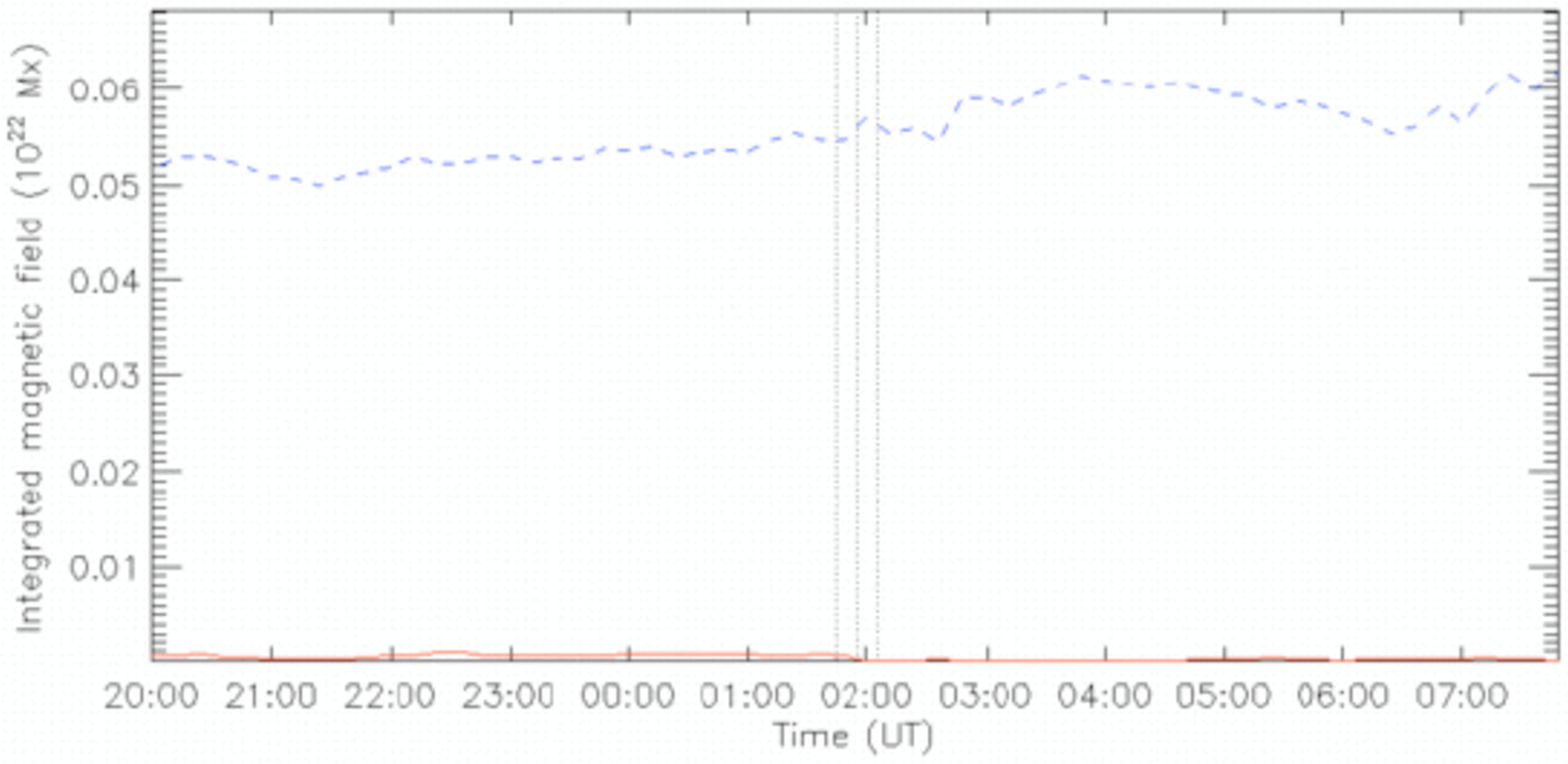}}
\resizebox{0.8\textwidth}{!}{\includegraphics*[20,250][600,550]{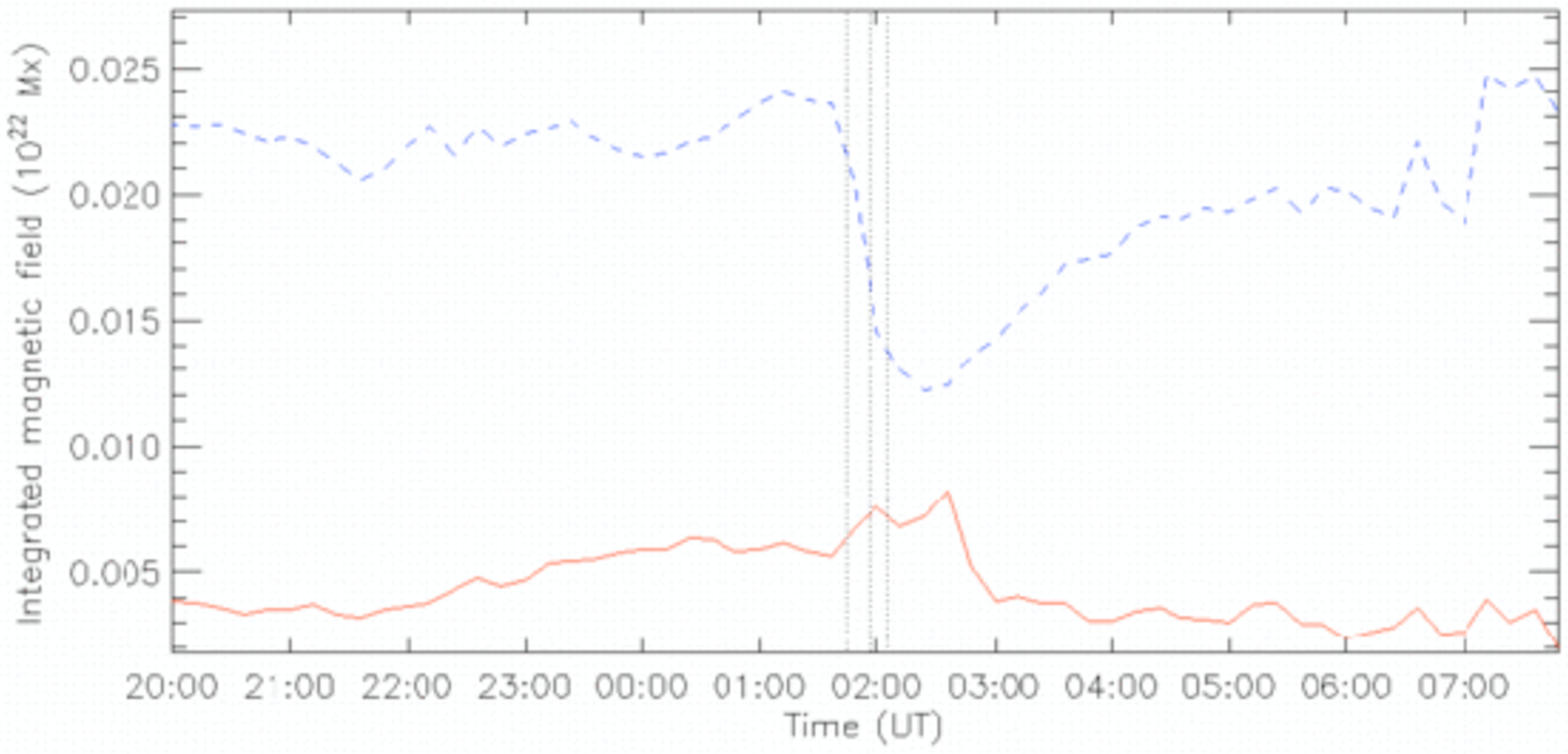}}
\resizebox{0.8\textwidth}{!}{\includegraphics*[20,250][600,550]{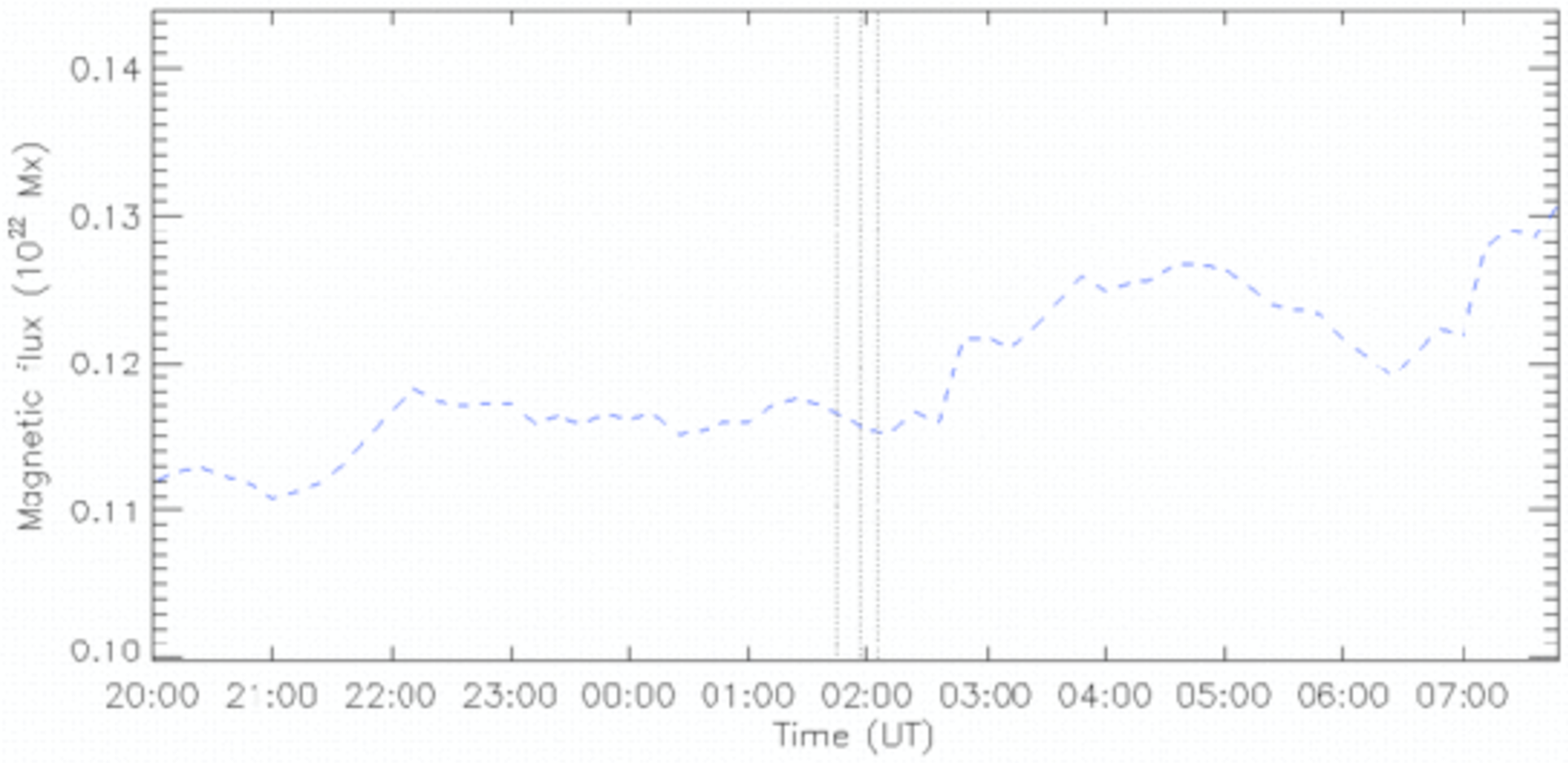}}
\end{center}
\caption{The radial integrated magnetic field $B_R^\mathrm{NS}$ (top), and the azimuthal, $B_{\theta}^\mathrm{NS}$ (middle), and vertical, $B_Z^\mathrm{NS}$ (bottom) integrated field components in the inner negative sunspot are plotted as functions of time. The red/blue solid/dashed lines represent positive/negative field. The vertical lines represent the GOES flare start, peak and end times.}
\label{fig:fint}
\end{figure}

\begin{figure} 
\begin{center}
\resizebox{0.8\textwidth}{!}{\includegraphics*[20,250][600,550]{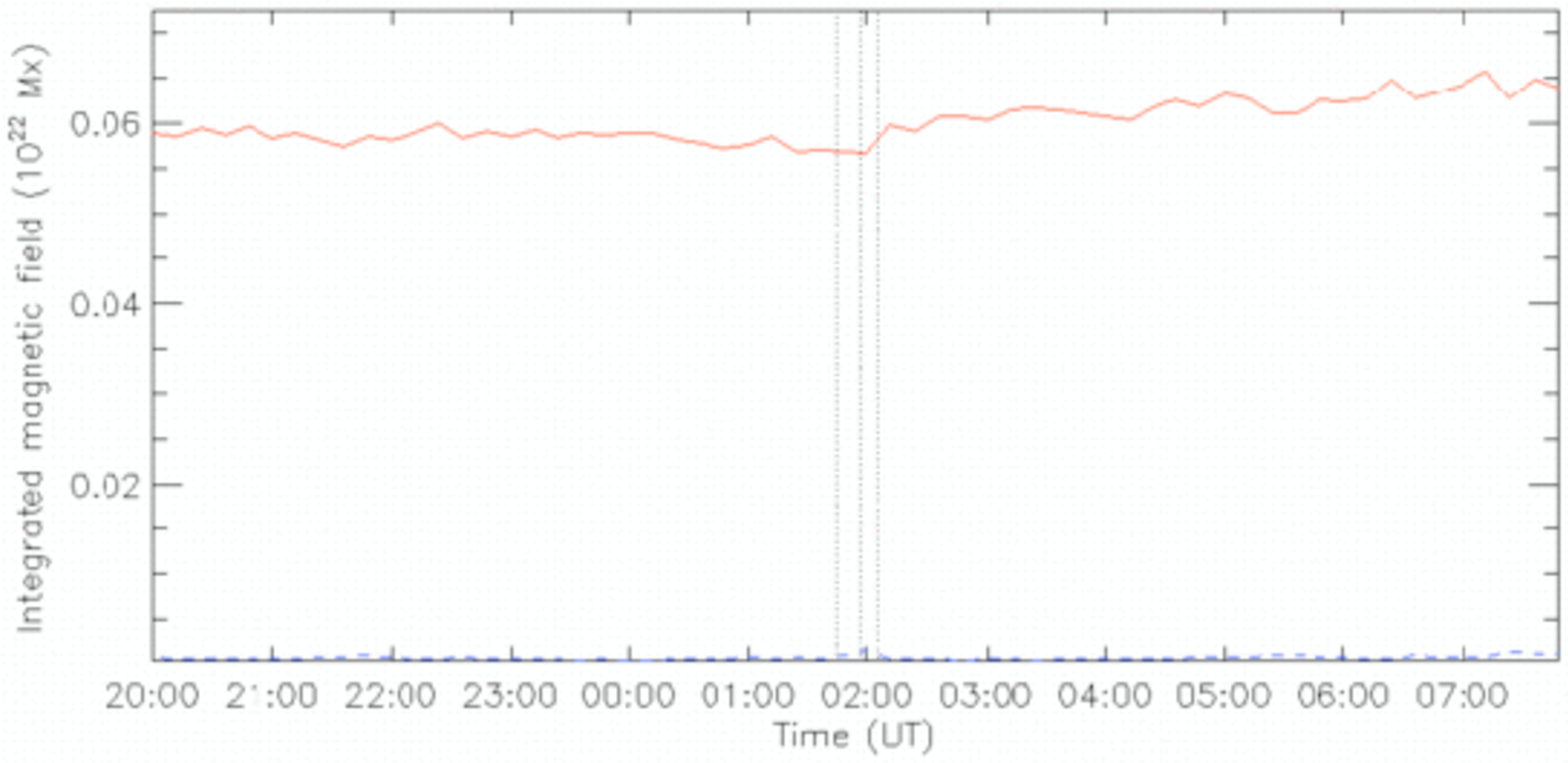}}
\resizebox{0.8\textwidth}{!}{\includegraphics*[20,250][600,550]{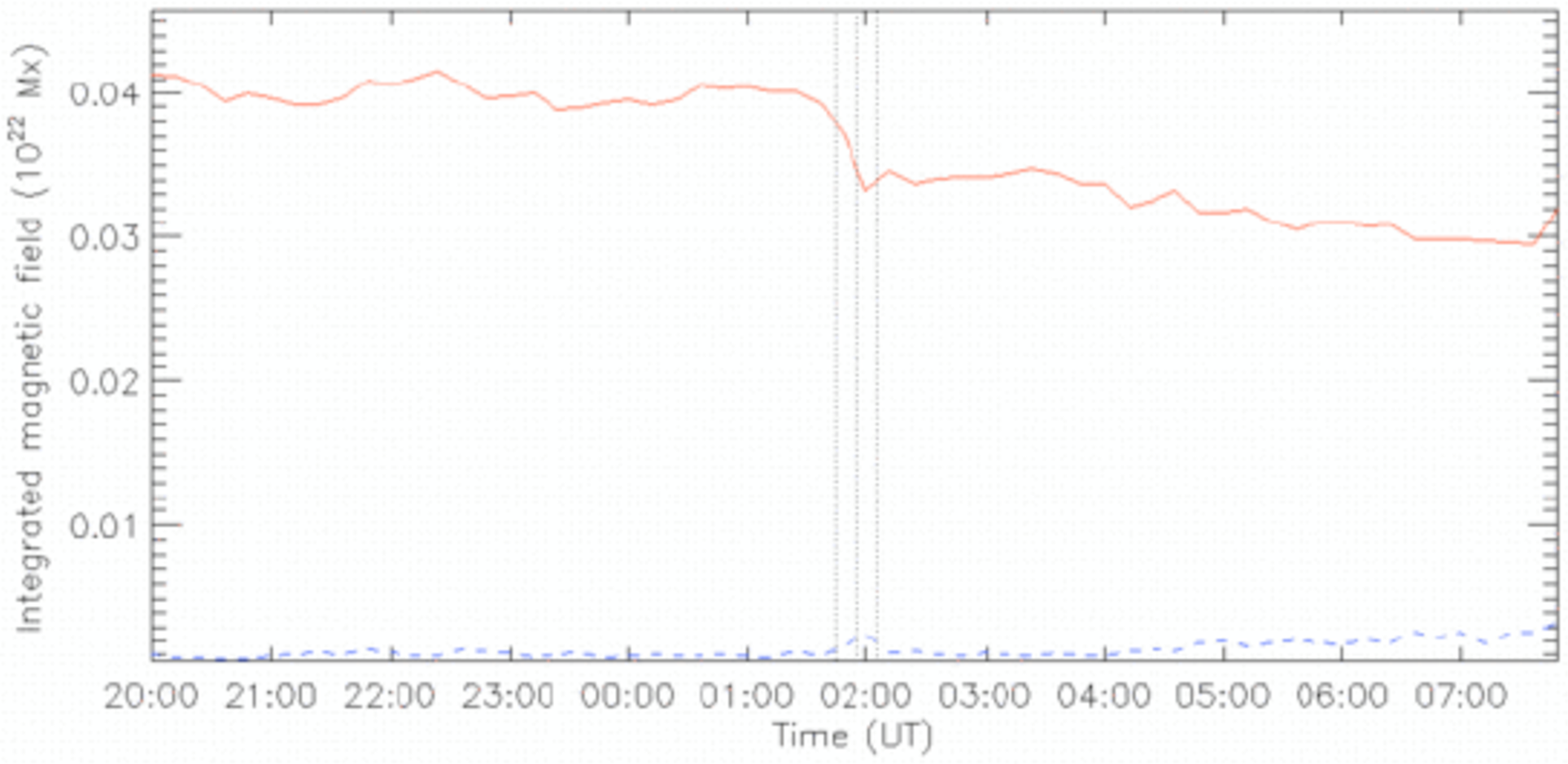}}
\resizebox{0.8\textwidth}{!}{\includegraphics*[20,250][600,550]{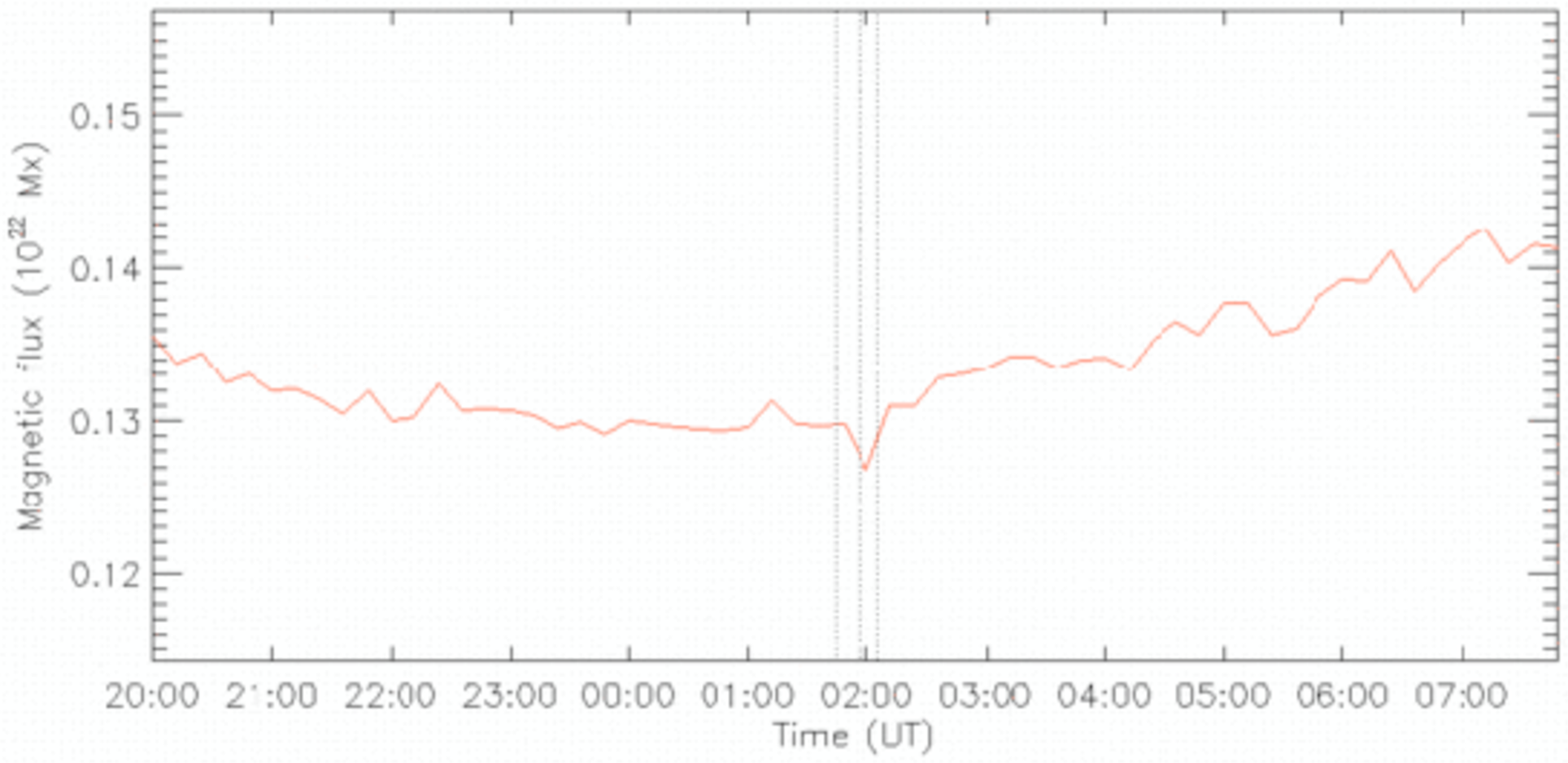}}
\end{center}
\caption{The radial integrated magnetic field $B_R^\mathrm{PS}$ (top), and the azimuthal, $B_{\theta}^\mathrm{PS}$ (middle), and vertical, $B_Z^\mathrm{PS}$ (bottom) integrated field components in the inner positive sunspot are plotted as functions of time. The red/blue solid/dashed lines represent positive/negative field. The vertical lines represent the GOES flare start, peak and end times.}
\label{fig:fipt}
\end{figure}

We next discuss the temporal profiles of the magnetic changes, shown in Figures \ref{fig:fnlt1}-\ref{fig:fipt}. These and subsequent plots of temporal changes were derived by calculating area integrals of the field components over chosen photospheric areas in the 60 12-minute images, from 20:00~UT on 14 February to 7:48~UT on 15 February. For example, the top panel of Figure~\ref{fig:fnlt1} shows the evolution of the integrated magnetic field strength

\begin{equation}
B^\mathrm{NL}=\int_{A_\mathrm{NL}} B\ \mathrm{d}A,
\label{eq:BNLint}
\end{equation}

\noindent near the neutral line between 20:00~UT on 14 February and 7:48~UT on 15 February. The area $A_\mathrm{NL}$ corresponds to the rectangular region near the neutral line marked in Figure~\ref{fig:brdbr}. The middle and bottom panels of Figure~\ref{fig:fnlt1} show the equivalent integrals $B_r^\mathrm{NL}$ and $B_\mathrm{h}^\mathrm{NL}$ of $B_r$ and $B_\mathrm{h}$. Near the neutral line, $B^\mathrm{NL}$ increased abruptly during the flare because of an increase in the horizontal field there. From the temporal plots it is clear that the vertical field near the neutral line did not change significantly overall during the flare, and the change that occurred did not have a permanent effect.

Figure~\ref{fig:fnlt2} shows the evolution of the integrated horizontal magnetic field components parallel and perpendicular to the neutral line, $B_{\parallel}^\mathrm{NL}$ and $B_{\perp}^\mathrm{NL}$. These directions are defined by the black rectangle in Figure~\ref{fig:brdbr} and the integrals are evaluated over the area $A_\mathrm{NL}$ represented by this rectangle. The parallel direction is the direction of the long edges of the rectangle, pointing approximately west-north-west. The perpendicular direction is the direction of the short edges of the rectangle, pointing approximately north-north-east. The horizontal field increased during the flare near the neutral line, both parallel and perpendicular to the neutral line. The change in the horizontal component parallel to the neutral line was the most significant change. The pre- and post-flare evolution of the horizontal field was more steady in the parallel than in the perpendicular component. The total field near the neutral line decreased steadily after the flare, reaching its pre-flare value about five hours after the flare. This decrease was due to changes in both horizontal and vertical components. The horizontal parallel component, however, remained significantly stronger five hours after the flare than its pre-flare value.

Flare-induced line profile changes can produce signatures that do not represent real changes in the magnetic field as discussed by Sudol and Harvey~(2005) - see the bottom part of their Figure~1. Working with GONG 1-minute longitudinal field images, Sudol and Harvey~(2005) and Petrie and Sudol~(2010) fitted a $\tan^{-1}$ step-like function to the time profile of each pixel, applied selection criteria based on the quality of the function fits, and inspected the results for representative pixels by eye. For the HMI vector data this approach is not as helpful because the sensitivity of the HMI vector field inversions is not as good as the sensitivity of the GONG data. Furthermore, the 12-minute time resolution does not resolve as many of the pixel field changes as the 1-minute GONG data do. Fits of $\tan^{-1}$ functions to HMI pixel time profiles are therefore not so useful in distinguising real field evolution from artifacts.

We have derived estimates of the field vector changes in two ways. We derived spatial maps by calculating pixel-by-pixel differences between before/after image pairs, the before image being composed of observations of the field before the published GOES start time of the X-ray flare and the after image deriving from observations of the field after the main flare-related field changes have taken place. This calculation therefore excluded the time when most of the flare emission transients are expected to have occurred and so the derived difference maps are not expected to be significantly compromised by such artifacts. In the second calculation we integrated field components over many-pixel regions of interest such as the rectangle surrounding the main neutral line marked in Figure~\ref{fig:brdbr}. We evaluated these integrals for all 60 magnetograms and formed time series of field integrals. This calculation does not avoid the time of the flare when emission might occur. We therefore need to be on the lookout for emission artifacts in the curves of the integrated quantities. Figures~\ref{fig:fnlt1} and \ref{fig:fnlt2} show that in this data set there is no significant signature of an emission artifact - compare with the bottom part of Sudol and Harvey's~(2005) Figure~1. In particular, the profiles of $B^\mathrm{NL}$, $B_\mathrm{h}^\mathrm{NL}$ and $B_{\parallel}^\mathrm{NL}$ have clear, stepwise changes with no sign of an emission artifact. This shows that, while emission transients may have affected some of the pixels, the calculations of the integrated quantities plotted in Figures~\ref{fig:fnlt1} and \ref{fig:fnlt2} were not significantly compromised by artifacts.

Besides occurring near neutral lines, abrupt field changes have been observed to occur in sunspots, as we discussed in Section~\ref{s:introduction}. To accompany the above analysis of the changes in heliographic coordinates, separate analyses of the field changes were performed for the sunspots of the active region in local cylindrical coordinates $(R,\Theta ,Z)$ with $R=0$ located at the sunspot center in each frame. The sunspot locations were tracked from frame to frame by first approximating them as linear functions of time and then searching for the location where the radial field vanished and the integral of the surrounding radial field was maximized. Smooth functions of position resulted from these estimates.

Figures~\ref{fig:fint} and \ref{fig:fipt} show the magnetic field evolution near the negative and positive inner sunspots in the radial, azimuthal and vertical directions as functions of time. Here the field components $B_i$ are integrated over the negative and positive sunspot areas $A^\mathrm{NS}$ and $A^\mathrm{PS}$ to give the integrated field components $B_i^\mathrm{NS}$ and $B_i^\mathrm{PS}$, where $i=R, \Theta, Z$. In the positive/negative sunspot all three integrated field components were positive/negative. A flux tube connecting the two spots would therefore be expected to have positive (right-handed) magnetic twist. Nonlinear force-free coronal field models extrapolated from these HMI photospheric vector magnetograms do indeed have positive relative magnetic helicity (Sun et al.~2012). The positive sunspot had nearly twice as much integrated azimuthal field as the negative sunspot. The other components were closer to equal strength in the two spots, with the vertical component larger than the other components in each case.  In the negative sunspot the dominant negative azimuthal component abruptly decreased during the flare and the positive azimuthal field continued a gradually increasing trend through the flare, so that the net integrated azimuthal field was small at the end of the flare. After the flare the dominant negative azimuthal component gradually returned to its original strength over the next several hours and the negative azimuthal field abruptly decreased, increasing the net integrated azimuthal field to almost its pre-flare value. The other two components did not show such a striking change. In the positive sunspot the dominant positive azimuthal component abruptly decreased during the flare and in this case the change was permanent. The negative integrated azimuthal field increased briefly during the flare but it was much smaller than the positive integrated azimuthal field throughout the time series. The other two components did not show such a striking change. In both spots the azimuthal field components changed more than the other components.

The magnetic twist of sunspot fields has been observed to decrease abruptly as a result of flares (Ravindra et al.~2011, Inoue et al.~2011). Modeling the emergence of a twisted flux tube from the interior through the photosphere into the corona, Longcope and Welsch~(2000) predicted that the expansion of the tube into the corona would redistribute the twist, creating an imbalance of torque at the photosphere-corona interface which would lead to a net rotation between the two photospheric footprints of the tube, reducing the coronal twist. The evolution of magnetic twist in emerging active regions observed by Pevtsov et al.~(2003) was found to be in agreement with Longcope and Welsch's~(2000) predictions. Magnetic helicity is not easily dissipated in the corona (Berger~1984) and is believed to accumulate there until bodily removed by coronal mass ejections (Low~2001). When twist is removed from coronal fields by coronal mass ejections, sub-photospheric fields could re-supply the twist until a new equilibrium is established, i.e., the rotation could be a reaction to the removal of twist from active region magnetic field (Pevtsov~2003, 2012). In our data the azimuthal field decreased abruptly in both sunspots during the flare, in agreement with this theoretical picture. Subsequently, the azimuthal field in the negative spot increased steadily, returning to its pre-flare value in a few hours. The positive spot had more azimuthal field than the negative spot throughout this series of observations, but showed no post-flare azimuthal field increase during the six hours of post-field-change observations analyzed here. It would be interesting to study the evolution of such fields over longer time intervals to see if the expected azimuthal field increases generally take place after flares.

We also computed spatial maps and temporal profiles of shear angles of the magnetic field. This is the angle between the observed horizontal photospheric field and the horizontal field of the unique potential field whose vertical component agrees with the observed vertical field distribution. This shear angle increased significantly as a result of the flare and the increases were mostly concentrated near the neutral line, following the pattern of the horizontal field changes. We do not show the plots here. The average shear angle near the neutral line as a function of time can be seen in Wang et al.~(2012). 

\section{The electric current}
\label{s:current}

\begin{figure} 
\begin{center}
\resizebox{0.99\textwidth}{!}{\includegraphics*[80,150][560,560]{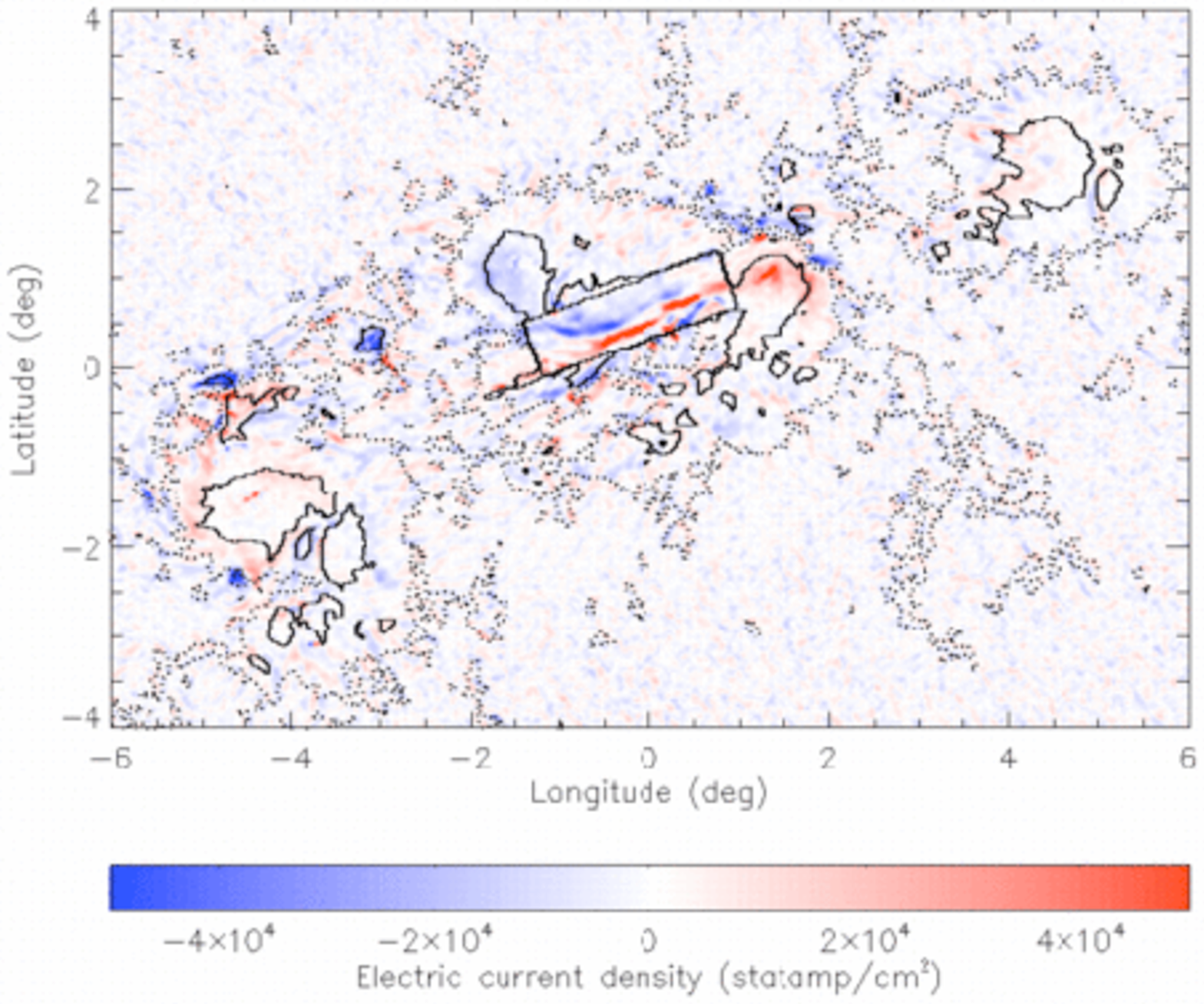}}
\resizebox{0.99\textwidth}{!}{\includegraphics*[80,170][560,560]{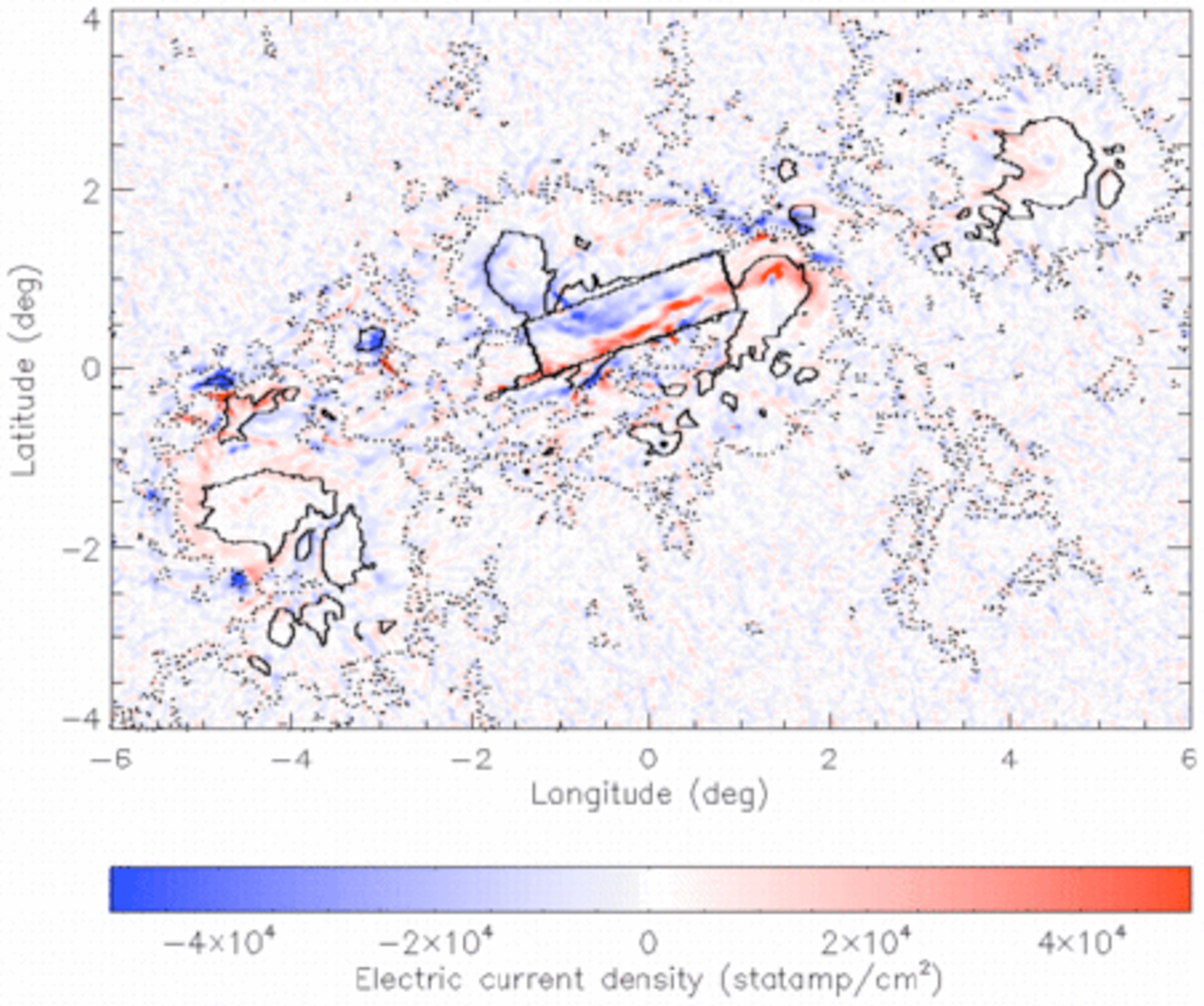}}
\end{center}
\caption{Shown are the vertical electric current density, $J_r$, before (top) and after (bottom) the main flare-related field changes. Red/blue coloring represents positive/negative vertical current with saturation values $5\times 10^4$~statamp\`ere/cm$^2$. The black rectangles mark the region near the neutral line that is used in the analysis.  The solid and dotted contours indicate strong ($\mathrm{|}B_r\mathrm{|}>1000$~G) and quite strong  ($\mathrm{|}B_r\mathrm{|}>100$~G) fields, respectively.}
\label{fig:jr}
\end{figure}

Plotted in Figure~\ref{fig:jr} are spatial maps of the vertical current density $J_r$ before and after the flare. Note the current reversal at the neutral line and the uniformity of sign on each side of the neutral line. One striking feature of Figure~\ref{fig:jr} is that in the two sunspots neighboring the neutral line the electric current is almost entirely of one polarity, positive/negative in the positive/negative sunspot. Analyzing 12 sunspots observed by \textit{Hinode}, by the \textit{Solar Optical Telescope/Spectro-polarimeter} instrument, Venkatakrishnan and Tiwari~(2009) found that net electric currents were generally absent from their data set. Our results clearly do not fit into this pattern. Such results have been reported in the past. Examining vector magnetic field data from 21 active regions observed by the \textit{Solar Magnetic Field Telescope} of the \textit{Huairou Solar Observing Station} of Beijing Astronomical Observatory, Wheatland~(2000) found that, while total active-region currents are well balanced, currents integrated over a given polarity of the magnetic field sum to quantities significantly different from zero, and so large-scale currents in active regions are typically unbalanced, implying that the magnetic field is not typically composed of isolated magnetic fibrils.

Comparing the two panels of Figure~\ref{fig:jr}, the radial current changes do not show an obvious pattern. One small but striking change is that the two sunspots near the neutral line had small opposite-polarity concentrations of current near their centers after the flare-related field changes. These were not present before the flare.


\begin{figure} 
\begin{center}
\resizebox{0.8\textwidth}{!}{\includegraphics*[10,250][600,550]{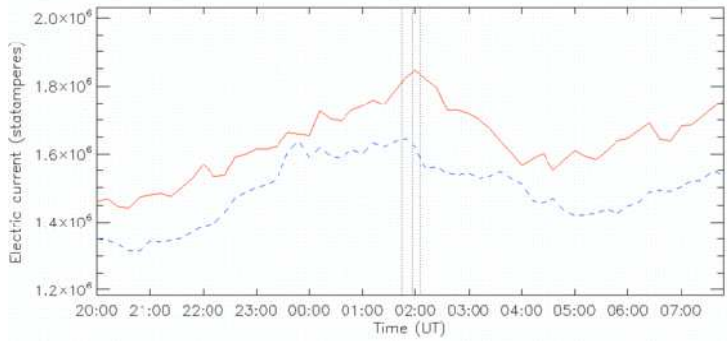}}
\end{center}
\caption{The vertical current density, $J_r$ near the neutral line as a function of time is plotted here. The red/blue solid/dashed lines represent positive/negative current. The vertical lines represent the GOES flare start, peak and end times. The area of integration for the neutral-line calculation is indicated by the black rectangle in Figure~\ref{fig:brdbr}. }
\label{fig:jbignlt}
\end{figure}

\begin{figure} 
\begin{center}
\end{center}
\caption{The vertical current density, $J_z$ at the inner negative (top) and positive (bottom) sunspots as functions of time are plotted here. The red/blue solid/dashed lines represent positive/negative current. The vertical lines represent the GOES flare start, peak and end times.}
\label{fig:jinpt}
\end{figure}

Figure~\ref{fig:jbignlt} shows the evolution of the integrated vertical current, $J_r^\textrm{NL}$ near the neutral line. According to Figure~\ref{fig:jbignlt} the negative vertical current near the neutral line decreased abruptly during the flare, although this change was not significantly greater than the background changes. The positive current continued its increasing trend through the start of the flare and began to decrease after the GOES peak time. The time of the flare marks a change from an increasing to an decreasing trend in electric current evolution near the neutral line. The total current of the region (not shown) did not change significantly at the time of the flare.

Figure~\ref{fig:jinpt} shows the electric currents in the positive and negative inner sunspots as functions of time. The electric currents of the two sunspots behaved quite differently. In the negative sunspot the dominant negative current abruptly decreased during and after the flare, but abruptly increased again to its pre-flare value less than an hour after the end of the flare. The dominant positive current in the positive sunspot showed no significant change during the flare but began a decreasing trend during the hours following the flare. In both sunspots the non-dominant polarity increased abruptly during the flare, because of the small opposite-polarity current concentrations appearing after the flare close to the centers of the sunspots in the lower panel of Figure~\ref{fig:jr}. This feature survived long after the flare in the negative sunspot but quickly disappeared from the positive sunspot after the flare.



\section{The Lorentz force changes}
\label{s:lorentzfch}

We use the results of Fisher et al.~(2012) to estimate the changes in the Lorentz force vector acting on the volume below the photosphere as a result of the flare. The Lorentz force per unit volume ${\bf f}_L$ can be written as,

\begin{equation}
{\bf f}_L = {\bf\nabla}\cdot{\bf T} = {\partial T_{ij}\over{\partial x_j }} ,
\label{eq:vollorentz}
\end{equation}

\noindent where the Maxwell stress tensor $[T_{ij}]$ in local Cartesian heliographic coordinates is,

\begin{equation}
T=\frac{1}{8\pi}\left[
\begin{array}{ccc}
B_r^2-B_{\theta}^2-B_{\phi}^2 & 2B_rB_{\theta} & 2B_rB_{\phi} \\
2B_rB_{\theta} & B_{\theta}^2-B_r^2-B_{\phi}^2 & 2B_{\theta}B_{\phi} \\
2B_rB_{\phi} & 2B_{\theta}B_{\phi} & B_{\phi}^2-B_r^2-B_{\theta}^2
\end{array}\right] .
\label{eq:maxwelltensor}
\end{equation}

Fisher et al.~(2012) evaluated the total Lorentz force over an atmospheric volume surrounding an isolated flaring active region by integrating Equation~(\ref{eq:vollorentz}) over this volume, whose lower boundary is identified with the photosphere, with upper boundary far above the photosphere, and side boundaries connecting these surfaces to form a closed volume $V$ as shown in Figure~1 of Fisher et al.~(2012). Evaluating the volume integral of Equation~(\ref{eq:vollorentz}) using Gauss's divergence theorem then gives (Fisher et al.~2012),

\begin{equation}
{\bf F}_L = \int_V {\bf\nabla}\cdot{\bf T}\ \mathrm{d}V = \int_{A_\mathrm{tot}} {\bf T}\cdot{\bf\hat{n}}\ \mathrm{d}A,
\label{eq:gaussdiv}
\end{equation}

\noindent where the area integral is evaluated over all surfaces of the volume, denoted by $A_\mathrm{tot}$, with unit normal vector ${\bf\hat{n}}$. As Fisher et al.~(2012)  argue, if the upper boundary of the volume is so far above the photosphere and the side boundaries are distant enough from the active region that the field integrals over these surfaces are negligible, then the surface integral of Equation~(\ref{eq:gaussdiv}) reduces to an integral over the photospheric lower boundary $A_\mathrm{ph}$ only. In this case, for the force acting on the volume below the photosphere, ${\bf\hat{n}} = {\bf\hat{r}}$ and ${\bf B}\cdot{\bf\hat{n}} = B_r$ and,

\begin{equation}
F_r = {1 \over 8 \pi} \int_{A_\mathrm{ph}} ( B_r^2 - B_\mathrm{h}^2 )\ \mathrm{d}A,
\label{eq:fr}
\end{equation}
and
\begin{equation}
{\bf F}_\mathrm{h} = {1 \over 4 \pi} \int_{A_\mathrm{ph}} ( B_r {\bf B}_\mathrm{h} )\ \mathrm{d}A.
\label{eq:fh}
\end{equation}

Then, assuming that the photospheric vector field is observed over a photospheric area $A_\mathrm{ph}$ at two times, $t=0$ before the field changes begin, and $t=\delta t$ after the main field changes have occurred, the corresponding changes in the Lorentz force vector components between these two times are given by Equations~(17) and (18) of Fisher et al.~(2012):

\begin{equation}
\delta F_r = {1 \over 8 \pi} \int_{A_\mathrm{ph}} ( \delta B_r^2 - \delta B_\mathrm{h}^2 )\ \mathrm{d}A,
\label{eq:deltafr}
\end{equation}
and
\begin{equation}
\delta {\bf F}_\mathrm{h} = {1 \over 4 \pi} \int_{A_\mathrm{ph}} \delta ( B_r {\bf B}_\mathrm{h} )\ \mathrm{d}A,
\label{eq:deltafh}
\end{equation}
where at a fixed location in the photosphere
\begin{eqnarray}
\delta B_\mathrm{h}^2 & = & B_\mathrm{h}^2(\delta t) -B_\mathrm{h}^2(0)\ ,\\
\delta B_r^2 & = & B_r^2(\delta t) -B_r^2(0)\ ,\\
\delta ( B_r {\bf B}_\mathrm{h} ) & = & B_r(\delta t){\bf B}_\mathrm{h}(\delta t) - B_r(0) {\bf B}_\mathrm{h}(0)\ .
\end{eqnarray}

\noindent The Lorentz force acting on the atmosphere above the photosphere is equal and opposite to the force acting on the volume at and below the photosphere (Fisher et al.~2012).

\begin{figure} 
\begin{center}
\resizebox{0.99\textwidth}{!}{\includegraphics*[80,150][560,560]{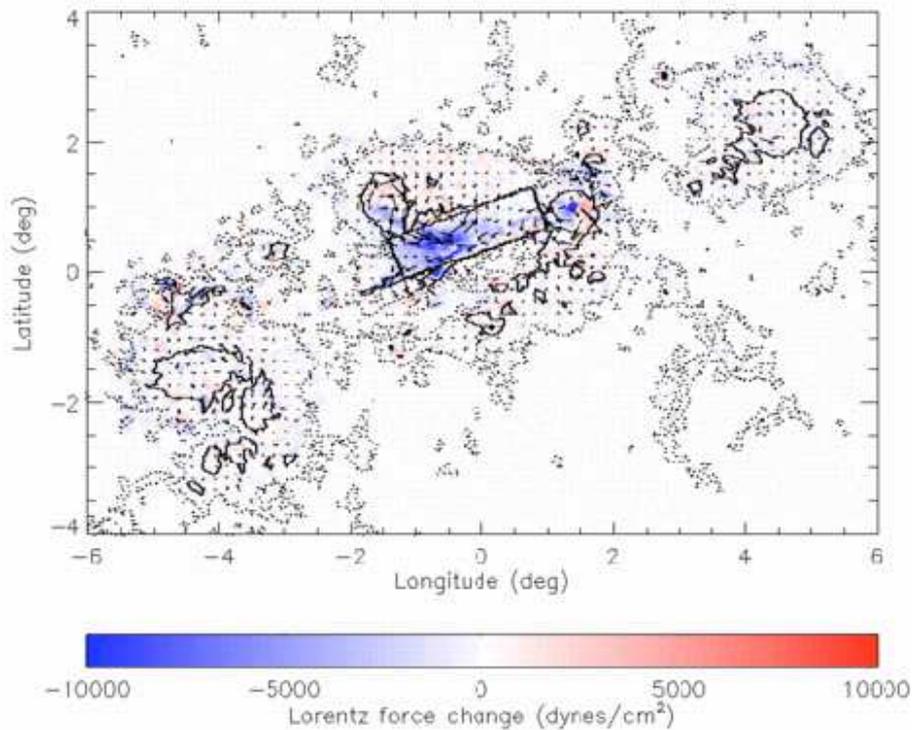}}
\end{center}
\caption{Shown are the Lorentz force vector changes during the flare. The vertical component, $\delta F_r$ is indicated by the color scale and the horizontal components by the arrows with saturation values $10^4$~dynes/cm$^2$ for the color scale and $2.5\times 10^3$~dynes/cm$^2$ for the arrows. Red/blue coloring represents positive/negative (upward/downward) Lorentz force change. The black rectangle marks the region near the neutral line that is used in the analysis.  The solid and dotted contours indicate strong ($\mathrm{|}B_r\mathrm{|}>1000$~G) and quite strong  ($\mathrm{|}B_r\mathrm{|}>100$~G) fields, respectively.}
\label{fig:dfr}
\end{figure}

Figure~\ref{fig:dfr} shows a spatial map of the Lorentz force changes $F_r$ and ${\bf F}_\mathrm{h}$, derived by evaluating the integrals in Equations~(\ref{eq:deltafr}) and (\ref{eq:deltafh}) pixel by pixel. The sums of the distributions shown in Figure~\ref{fig:dfr} over the entire photospheric area gives the estimate for the total Lorentz force vector described by Fisher et al.~(2012). We argue that Figure~\ref{fig:dfr} also gives a useful estimate of the spatial distributions of the Lorentz force vector components across the region.

According to Equation~(\ref{eq:vollorentz}) the divergence of the Maxwell stress tensor $\bf T$ gives the Lorentz force density. Because the coronal plasma is likely to be an approximately force-free medium and the photospheric plasma is not, we expect $\bf T$ to be nearly divergence-free everywhere in the volume of integration $V$ in Equation~(\ref{eq:gaussdiv}) except at the lower boundary, leading us to expect most contributions to the volume integral in Equation~(\ref{eq:gaussdiv}) to come from the lower boundary. We also expect the field components themselves to be stronger at the lower boundary than elsewhere in the volume because photospheric active region fields are significantly stronger than their coronal counterparts.

Evaluating the volume integral of Equation~(\ref{eq:gaussdiv}) over a single pixel, the volume of integration becomes a very tall, thin pillbox-shaped volume extending high into the atmosphere, whose photospheric footprint coincides with the single pixel. The contribution from the lower boundary is given by Equations~(\ref{eq:fr}) and (\ref{eq:fh}). The contribution from the upper boundary is negligible, provided that this boundary is sufficiently high above the photosphere, but the contributions from the side boundaries need to be accounted for. For example, from Equations~(\ref{eq:maxwelltensor}) and (\ref{eq:gaussdiv}), the contributions to $F_{\theta}$ from the walls with normal unit vectors $\pm\bf\hat{\theta}$ are,

\begin{equation}
F_{\theta}^{(\pm\bf\hat{\theta})} = \pm\frac{1}{8\pi} \int_{A_{\pm\bf\hat{\theta}}} B_{\theta}^2-B_r^2-B_{\phi}^2\ \mathrm{d}A.
\label{thetacont}
\end{equation}

\noindent The contributions to $F_{\theta}$ from the other boundaries can be readily be derived from Equations~(\ref{eq:maxwelltensor}) and (\ref{eq:gaussdiv}). These contributions involve coronal fields that are generally significantly weaker than their photospheric counterparts on the lower boundary. Furthermore, because the field distributions for major, well-resolved magnetic structures are likely to be similar on the two opposing side boundaries, situated a pixel-length apart, the contributions to Equation~(\ref{eq:gaussdiv}) are expected to approximately cancel each other and so their contribution can be neglected. For photospheric field structures composed of many pixels whose field patterns are well resolved, such as the region around the main neutral line and the major sunspots shown in Figure~\ref{fig:brdbr}, the errors in the estimated Lorentz force density changes shown in Figure~\ref{fig:dfr} are expected to be small. Also the sum of these changes exactly matches the area integral described by Fisher et al.~(2012).

In the example shown in Figure~\ref{fig:dfr}, the horizontal field changes $\delta{\bf B}_\mathrm{h}$ near the main neutral line are increases in horizontal field strength, $\delta B_\mathrm{h}^2>0$, and are significantly greater than the vertical field changes $\delta B_r^2$ (compare the middle and bottom panels of Figure~\ref{fig:fnlt1}). Equation~(\ref{eq:deltafr}) therefore leads us to expect the vertical Lorentz force change to have been predominantly downward there. Figure~\ref{fig:dfr} shows the spatial distribution of the change in the Lorentz force components during the flare. As expected, near the main neutral line the Lorentz force clearly acted downwards into the photosphere. This behavior was anticipated to occur near neutral lines of flaring active regions by Hudson, Fisher and Welsch (2008) and Fisher et al.~(2012), and has been found in past estimates of Lorentz force changes by Wang and Liu~(2010) and Petrie and Sudol~(2010). The two sunspots neighboring the neutral line appear to have undergone forces consistent with tilting motions towards the neutral line: their vertical force changes close to the neutral line were downward while those further from the neutral line were upward. Some evidence of corresponding field changes can be seen in Figure~\ref{fig:brdbr} where in these spots some vertical fields close to the neutral line became weaker and some far from the neutral line became stronger. However the signature in Figure~\ref{fig:dfr} is clearer.

The horizontal Lorentz force changes also show clear patterns. Equation~\ref{eq:deltafh} implies that, wherever the vertical field does not change significantly compared to the horizontal changes and is positive/negative, the horizontal Lorentz force changes $\delta{\bf F}_\mathrm{h}$ should be parallel/anti-parallel to the horizontal field changes $\delta {\bf B}_\mathrm{h}$. We already know from Figure~\ref{fig:brdbr} that on both sides of the neutral line $\delta {\bf B}_\mathrm{h}$ pointed eastward and approximately parallel to the neutral line. Figure~\ref{fig:dfr} shows that the horizontal Lorentz force change $\delta{\bf F}_\mathrm{h}$ acted in opposite directions along each side of the neutral line, with the changes on the southern positive side pointing eastward and those on the northern negative side westward as expected. Since the sheared field at the neutral line pointed eastward, these Lorentz force changes are consistent with a reduction of the magnetic shear parallel to the neutral line. These horizontal Lorentz force changes were directed against the shear flow pattern described by Beauregard, Verma and Denker~(2012). It seems that the steady shear flow pattern created, or at least strengthened, the magnetic shear whereas the horizontal Lorentz force changes acted towards relaxing the shear. However, we know from Section~\ref{s:magch} that the parallel field component increased significantly during the flare. The horizontal Lorentz force pattern is a signature of the field contracting across the neutral line during the flare, tugging the photospheric fields on the two sides towards each other. The horizontal field that was added during the flare was associated with the large downward Lorentz force change described above, and was caused by sheared field collapsing towards the photospheric neutral line from above. This suggests that the field near the neutral line underwent a contraction during the flare, both vertically towards the photosphere and horizontally along the neutral line. The vertical collapse had the dominant effect on the field changes near the neutral line.

The horizontal force change patterns in the two sunspots near the neutral line are also striking. In both spots the horizontal changes had a dominant clockwise azimuthal component. These changes merged with the horizontal changes about the neutral line in acting westward on the northern side and eastward on the southern side of each spot. Because the vertical changes are smaller than the horizontal changes in Figures~\ref{fig:fint} and \ref{fig:fipt}, from Equation~(\ref{eq:deltafh}) we expect the Lorentz force change $\delta{\bf F}_\mathrm{h}$ to be parallel/anti-parallel to the horizontal field changes $\delta{\bf B}_\mathrm{h}$ and this is indeed the case. As Figure~\ref{fig:brdbr} shows, the positive/negative spot showed clockwise/anti-clockwise horizontal field changes, and the horizontal Lorentz force changes were clockwise in both spots as seen in Figure~\ref{fig:dfr}. Recall that the dominant azimuthal field changes were anti-clockwise in the negative spot and clockwise in the positive spot, reducing the azimuthal field in each spot. The clockwise Lorentz force changes in both spots are signatures of the azimuthal field being abruptly removed from the spots from above. Jiang et al.~(2012) detected clear clockwise rotation in the sunspot proper motions and associated these rotations with the development of the positive-helicity spiral pattern of the positive spot's penumbral filaments and the shearing of the main neutral line. The horizontal Lorentz force vector changes that we have derived from the vector field changes are therefore directed in the same azimuthal direction as these proper motions but, being applied from above, they relaxed the field instead of twisting it.

\begin{figure} 
\begin{center}
\resizebox{0.8\textwidth}{!}{\includegraphics*[20,250][600,550]{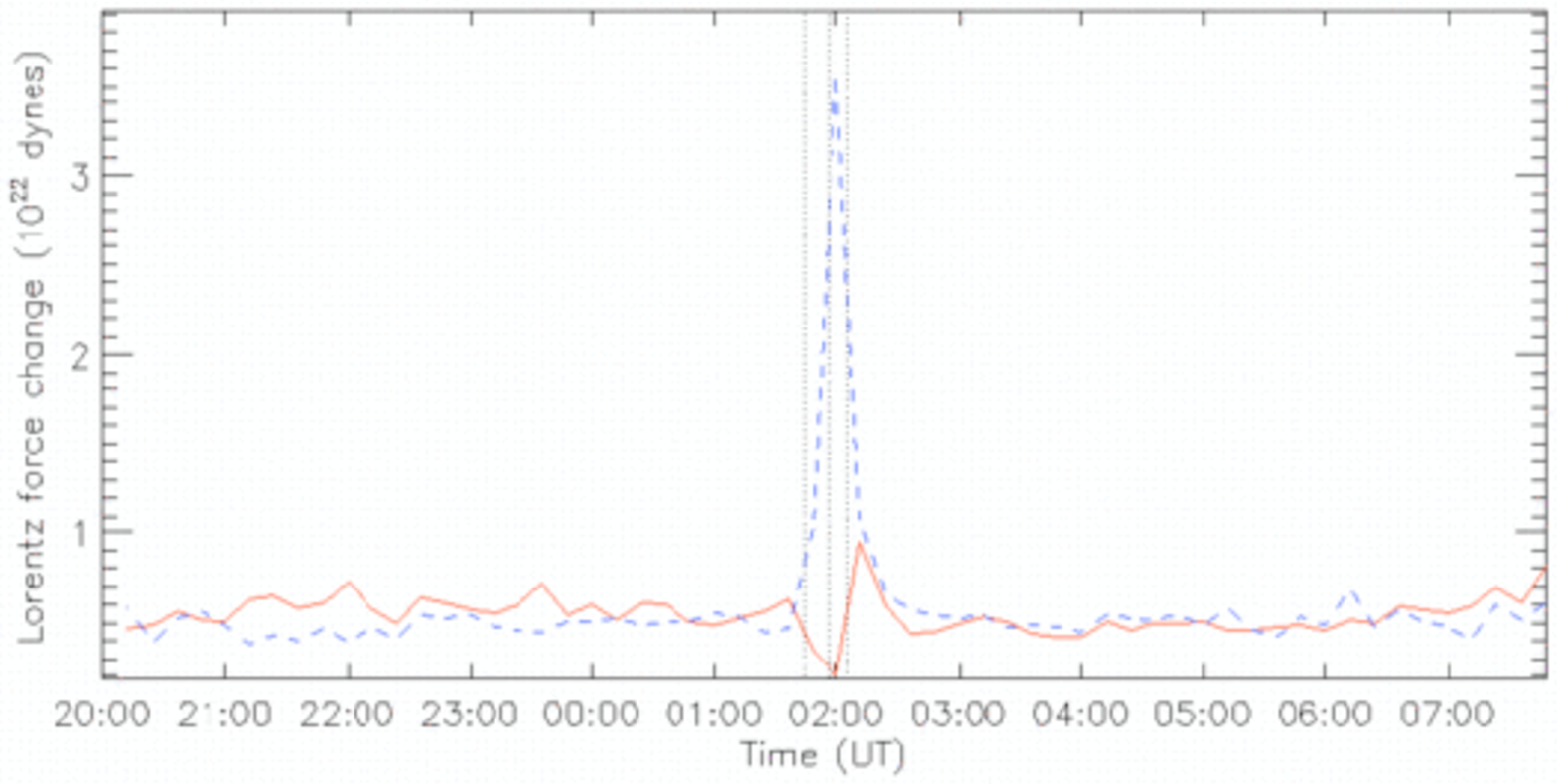}}
\resizebox{0.8\textwidth}{!}{\includegraphics*[20,250][600,550]{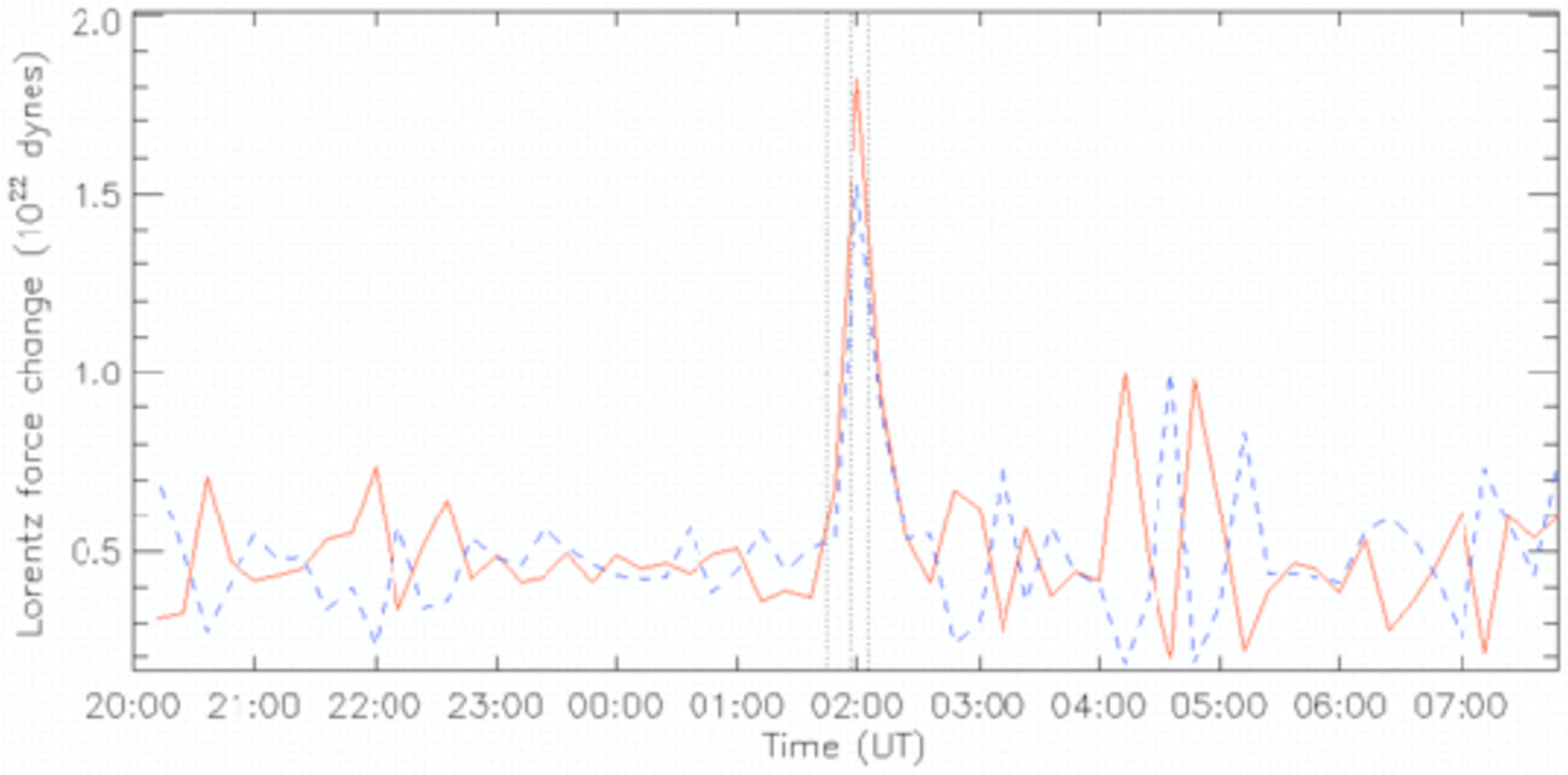}}
\resizebox{0.8\textwidth}{!}{\includegraphics*[20,250][600,550]{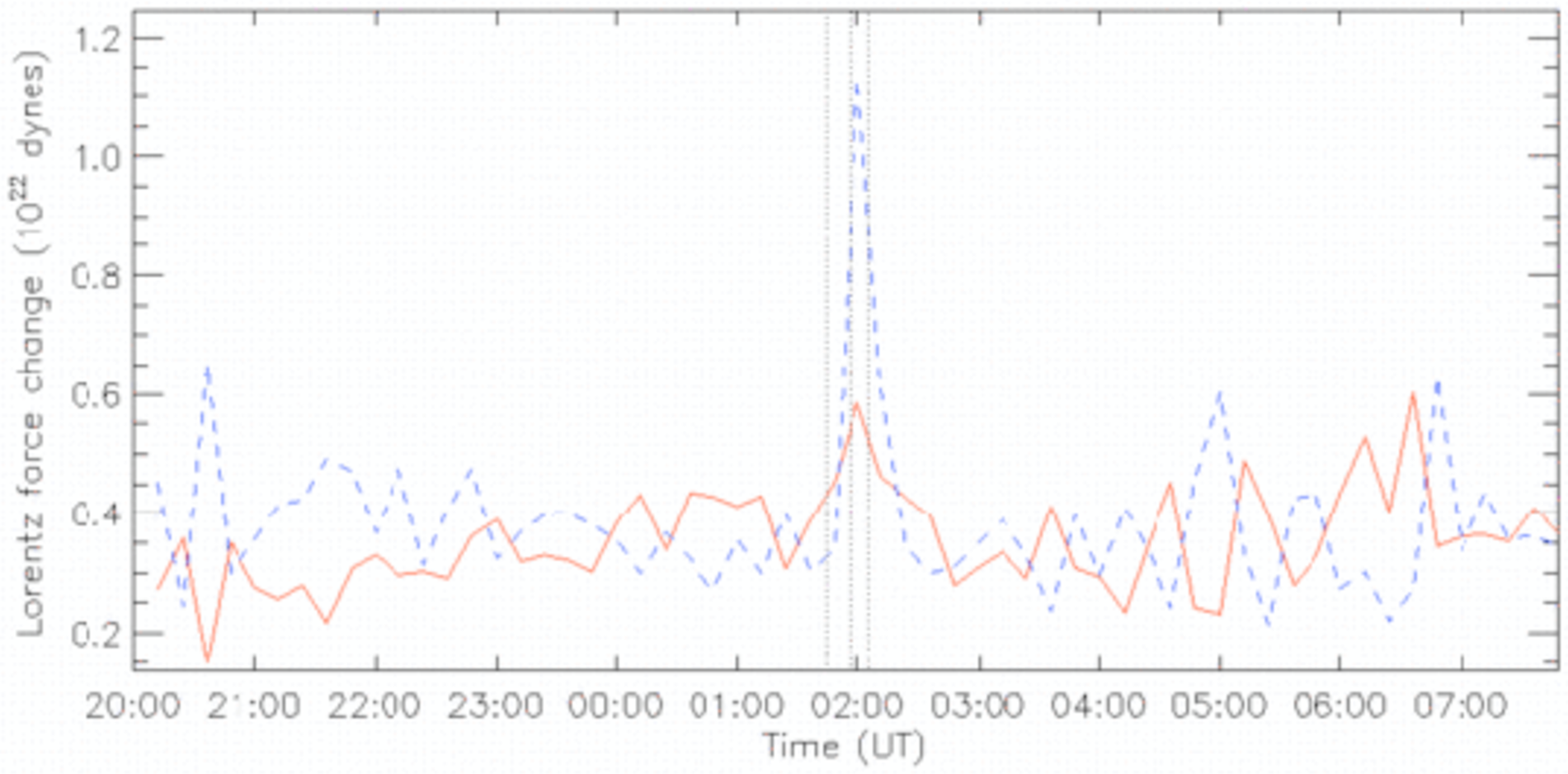}}
\end{center}
\caption{The Lorentz force vector components in the vertical ($\delta F_r^\mathrm{NL}$, top) and horizontal directions parallel ($\delta F_{\parallel}^\mathrm{NL}$, middle) and perpendicular ($\delta F_{\perp}^\mathrm{NL}$, bottom) to the neutral line are plotted as functions of time. The area of integration is indicated by the rectangle in Figure~\ref{fig:brdbr}. The red/blue solid/dashed lines represent positive/negative force changes. The vertical lines represent the GOES flare start, peak and end times.}
\label{fig:dfnlt}
\end{figure}

\begin{figure} 
\begin{center}
\end{center}
\caption{The Lorentz force vector changes near the negative inner sunspot in the radial ($\delta F_R^\mathrm{NS}$, top), azimuthal ($\delta F_{\Theta}^\mathrm{NS}$, middle) and vertical ($\delta F_Z^\mathrm{NS}$, bottom) directions as functions of time. The red/blue solid/dashed lines represent positive/negative force changes. The vertical lines represent the GOES flare start, peak and end times.}
\label{fig:dfint}
\end{figure}

\begin{figure} 
\begin{center}
\end{center}
\caption{The Lorentz force vector changes near the positive inner sunspot in the radial ($\delta F_R^\mathrm{PS}$, top), azimuthal ($\delta F_{\Theta}^\mathrm{PS}$, middle) and vertical ($\delta F_Z^\mathrm{pS}$, bottom) directions as functions of time. The red/blue solid/dashed lines represent positive/negative force changes. The vertical lines represent the GOES flare start, peak and end times.}
\label{fig:dfipt}
\end{figure}

Figure~\ref{fig:dfnlt} shows the Lorentz force vector changes in the vertical and horizontal directions parallel and perpendicular to the neutral line as functions of time. These plots show the Lorentz force change components in the vertical, horizontal parallel and horizontal perpendicular directions, $\delta F_R^\mathrm{NL}$, $\delta F_\parallel^\mathrm{NL}$ and $\delta F_\perp^\mathrm{NL}$, integrated over the area $A_\mathrm{NL}$ represented by the black rectangle shown in Figure~\ref{fig:brdbr}. The parallel direction is the direction of the long edges of the rectangle, pointing approximately west-north-west. The perpendicular direction is the direction of the short edges of the rectangle, pointing approximately north-north-east. These plots show sharp signatures of the abrupt magnetic changes during the flare in all three directions, particularly the very large downward changes and the changes in both directions, of almost equal total size, parallel to the neutral line. The sizes of these force changes, about $3.5\times 10^{22}$~dynes downward near the neutral line and almost as much in the horizontal directions, is larger than those found in the previous estimates of flare-related Lorentz force changes by Wang and Liu~(2010) and Petrie and Sudol~(2010). For the 2002 July 26 M8.7 flare Wang and Liu~(2010) found a downward force change of $1.6\times 10^{22}$~dynes. Petrie and Sudol~(2010) found a range of longitudinal force change estimates up to about $2\times 10^{22}$~dynes. Petrie and Sudol's estimates are likely to have been underestimates because they included only information on the longitudinal field component.

Figures~\ref{fig:dfipt} and \ref{fig:dfint} show the Lorentz force vector changes $\delta F_i^\mathrm{NS}$ and $\delta F_i^\mathrm{PS}$, where $i=R, \Theta , Z$, integrated over the negative and positive inner sunspot  areas $A_\mathrm{NS}$ and $A_\mathrm{PS}$ in the radial, azimuthal and vertical directions as functions of time. In both cases the main effect of the flare is a large negative azimuthal (clockwise) force change at the time of the flare, consistent with the arrows in Figure~\ref{fig:dfr}. Sizable force changes in other components are also evident in the plots, but these are not as large and are of mixed sign.

\section{Conclusion}
\label{s:conclusion}

We have analyzed in detail 12 hours of 12-minute SDO/HMI vector field observations covering the first X-class flare in Cycle 24, the X2.2 flare at 01:44 UT on 15 February 2011. This data set has given us the first opportunity to resolve spatial and temporal changes of field direction and strength, and their associated Lorentz force changes, in three spatial dimensions.

The main conclusions are:

\begin{enumerate}

\item Near the neutral line, the photospheric field vectors became stronger and more horizontal during the flare. This was due to an increase in strength of the horizontal field components near the neutral line. The increase in strength was most significant in the horizontal component parallel to the neutral line but the component perpendicular to the neutral line also increased in strength. The result was an increase in the shear of the field near the neutral line.

\item Perhaps surprisingly, the vertical field component did not show a significant, permanent overall change at the neutral line to compensate for the strengthened horizontal field. Instead, the increase in field at the neutral line was accompanied by a compensating field decrease in the surrounding volume. The total photospheric field of the active region did not change significantly during the flare.

\item The two sunspots near the main neutral line also showed significant field changes. In both cases the azimuthal field abruptly decreased during the flare but this change was permanent in only one of the spots.

\item The vertical electric current density near the main neutral line steadily increased until the time of the flare, then steadily decreased for a few hours after the flare.

\item The vertical Lorentz force had a large, abrupt downward change during the flare. This is consistent with past observations and with recent theoretical work.

\item The horizontal Lorentz force acted in opposite directions on each side of neutral line during the flare. The two sunspots at each end of the neutral line underwent abrupt torsional Lorentz force changes that merged with the shearing pattern of the neutral-line force changes and were consistent with the relaxation of twist. The shearing forces were consistent with a contraction of the field and a decrease of shear near the neutral line, whereas the field itself became more sheared as a result of the field collapsing towards the neutral line from the surrounding volume.

\end{enumerate}

Increased magnetic field tilts during flares at neutral lines have been detected many times in the past in vector measurements (Wang and Liu~2010, Wang et al.~2012) and also in the statistics of longitudinal measurements (Petrie and Sudol~2010). The HMI vector data have enabled us to provide spatial maps of these changes, allowing us to show that the changes of field tilt are not the result of a simple rotation of the magnetic vector towards the neutral line but a transfer of magnetic field towards the photospheric magnetic neutral line from the surrounding volume. The associated Lorentz force changes are also not consistent with a rotation of the magnetic vector towards the neutral line. The shearing pattern of the horizontal Lorentz forces and the related azimuthal forces in the neighboring sunspots would by themselves have reduced the shear of the field near the neutral line. However they were accompanied by a strong downward force change associated with the field collapsing downward from the surrounding volume. It was this process that was decisive in increasing the shear of the field around the neutral line. Wang~(2006) found from a study of high-cadence longitudinal magnetograms that some flares produced a decrease in magnetic shear along the main neutral line, while in other cases, the shear increased. If the two patterns of Lorentz force change, horizontal shearing patterns that act to decrease the shear of the photospheric field and downward forces that increase the shear, generally occur during flares, then this might imply that the vertical forces dominate during some flares and the horizontal forces dominate during others. Whether the collapsing field is highly sheared or not seems also to be an important factor in determining the outcome.  Nonlinear force-free field extrapolations have suggested that if the photospheric field shear increases during a flare, that the increase is localized at low heights and the shear decreases above a certain height (Jing et al.~2008, Sun et al.~2011, Liu et al.~2012). This is consistent with a sheared structure collapsing towards the neutral line, leaving a void above that is filled by more relaxed field.

In the future this work will be extended by studying many more examples and deducing which general patterns tend to dominate during flares. We are not currently capable of predicting the behavior of active regions and the occurrence of flares and ejections. Only with a comprehensive and detailed study of the governing fields will this become possible. This work has suggested that horizontal and vertical Lorentz force changes can have different effects on the shear of magnetic neutral-line fields, with the vertical changes dominating in the case of the X2.2 flare at 01:44 UT on 15 February 2011. In Petrie and Sudol's~(2010) statistical study of GONG 1-minute longitudinal magnetograms covering 77 major flares the horizontal field changes (the longitudinal field changes observed near the limb) were larger than the vertical field changes (the longitudinal field changes observed near disk-center), were usually negative on all parts of the disk investigated, and most of the derived Lorentz force changes, particularly the largest ones, pointed downward. These results suggest that the horizontal field changes and associated vertical Lorentz force changes tend to be, but are not always, the more important. By studying many more high-cadence vector-field measurements from HMI and the National Solar Observatory's \textit{Synoptic Optical Long-term Investigations of the Sun} (SOLIS) telescope, we might be able to find out what the factor is that determines which force changes are dominant, and whether this factor is related to the flare productivity of an active region. For example, if the horizontal motions are more significant, will more twist be injected from below as predicted by Longcope and Welsch~(2003), and will the region tend to flare again sooner? The relationship between sunspot twist and neutral-line shear is clearly non-trivial and will be investigated further. Sunspot twist, flux emergence and cancellation, and Hudson implosion all affect the distribution of Maxwell stresses in active-region magnetic fields in different ways, and a focused, observationally-driven study of the interplay of these processes will reveal much of the basic dynamics of flares.

%

%
\begin{acks}
I thank the referee for helpful comments that resulted in a much improved, clearer paper. SDO is a mission for NASA's Living With a Star program. I thank Sanjay Gosain and Alexei Pevtsov for discussions. This work was supported by NSF Award No. 106205 to the National Solar Observatory. 
\end{acks}

%

%
\bibliographystyle{spr-mp-sola-cnd} 


\IfFileExists{\jobname.bbl}{} {\typeout{}
\typeout{****************************************************}
\typeout{****************************************************}
\typeout{** Please run "bibtex \jobname" to obtain} \typeout{**
the bibliography and then re-run LaTeX} \typeout{** twice to fix
the references !}
\typeout{****************************************************}
\typeout{****************************************************}
\typeout{}}

\end{article} 
\end{document}